\documentclass[aps,prd,preprintnumbers,nofootinbib,twocolumn,superscriptaddress]{revtex4-1}

\usepackage{amsmath}
\usepackage{amssymb}
\usepackage{amsfonts}
\usepackage{mathrsfs}
\usepackage{dsfont}
\usepackage{slashed}
\usepackage{graphics}
\usepackage{bbm,bm}
\usepackage{psfrag}
\usepackage{epsfig}
\usepackage{booktabs}
\usepackage{epstopdf}
\usepackage{hyperref}
\usepackage{color}
\usepackage{stackengine}
\graphicspath{{figures/}}

\usepackage[utf8]{inputenc}

\usepackage[english]{babel}

\newcommand{\Eqref}[1]{Eq.~\eqref{#1}}
\newcommand{\Appref}[1]{App.~\ref{#1}}
\newcommand{\Secref}[1]{Sec.~\ref{#1}}
\newcommand{\Figref}[1]{Fig.~\ref{#1}}
\newcommand{\be}{\begin{eqnarray}}
\newcommand{\ee}{\end{eqnarray}}
\newcommand{\bna}{\begin{align}}
\newcommand{\ena}{\end{align}}
\newcommand{\bs}{\begin{split}}
\newcommand{\es}{\end{split}}
\newcommand{\widesim}[2][1.5]{
	\mathrel{\underset{#2}{\scalebox{#1}[1]{$\sim$}}}
}

\newcommand{\dd}{\mathrm{d}}
\newcommand{\DD}{\mathcal{D}}

\newcommand{\I}{\mathrm{i}}

\newcommand{\cG}{\mathcal{G}}
\newcommand{\cJ}{\mathcal{J}}
\newcommand{\cI}{\mathcal{I}}

\newcommand{\Det}{\mathrm{Det}}

\newcommand{\bC}{\bar{c}}
\newcommand{\C}{c}
\newcommand{\B}{b}
\newcommand{\no}{n}
\newcommand{\F}{\mathsf{F}}
\newcommand{\LL}{\mathsf{L}}
\newcommand{\X}{\mathsf{Q}}
\newcommand{\m}{\bar{m}}
\newcommand{\mm}{\bar{m}_\gh}
\newcommand{\bth}{\bar{\theta}}

\newcommand{\YM}{\mathrm{YM}}
\newcommand{\NL}{\mathrm{NL}}
\newcommand{\FP}{\mathrm{FP}}
\newcommand{\gf}{\mathrm{gf}}
\newcommand{\gh}{\mathrm{gh}}
\newcommand{\tot}{\mathrm{tot}}
\newcommand{\so}{\mathrm{so}}
\newcommand{\brs}{\mathrm{brst}}
\newcommand{\T}{\mathrm{T}}
\newcommand{\Lo}{\mathrm{L}}
\newcommand{\Tr}{\mathrm{Tr}}
\newcommand{\tGamma}{\widetilde{\Gamma}}
\newcommand{\G}{\mathcal{G}}

\begin{document}

\title{BRST invariant RG flows}

\author{Shimasadat Asnafi}
\email{shimaasnafi@gmail.com}
\affiliation{Department of Physics, Sharif University of Technology, P.O.~Box 11155-9161, Tehran-Iran}
\affiliation{School of Particles and Accelerators, Institute for Research in Fundamental Sciences (IPM) P.O.~Box 19395-5531, Tehran, Iran}
\affiliation{\mbox{\it Theoretisch-Physikalisches Institut, Friedrich-Schiller-Universit{\"a}t Jena,}
\mbox{\it D-07743 Jena, Germany}}
\affiliation{Helmholtz-Institut Jena, Fr\"obelstieg 3, D-07743 Jena, Germany}

\author{Holger Gies}
\email{holger.gies@uni-jena.de}
\affiliation{\mbox{\it Theoretisch-Physikalisches Institut, Friedrich-Schiller-Universit{\"a}t Jena,}
\mbox{\it D-07743 Jena, Germany}}
\affiliation{\mbox{\it Abbe Center of Photonics, Friedrich-Schiller-Universit{\"a}t Jena,}
	\mbox{\it D-07743 Jena, Germany}}
\affiliation{Helmholtz-Institut Jena, Fr\"obelstieg 3, D-07743 Jena, Germany}

\author{Luca Zambelli}
\email{luca.zambelli@uni-jena.de}
\affiliation{\mbox{\it Theoretisch-Physikalisches Institut, Friedrich-Schiller-Universit{\"a}t Jena,}
\mbox{\it D-07743 Jena, Germany}}


\begin{abstract}

A mass parameter for the gauge bosons in gauge-fixed four-dimensional Yang-Mills
theory can be accommodated in a local and manifestly BRST-invariant action.
The construction is based on the Faddeev-Popov method
involving a nonlinear gauge-fixing and a background Nakanishi-Lautrup field.
When applied to momentum-dependent masslike deformations,
this formalism leads to a full regularization of the theory which explicitly preserves
BRST symmetry. We deduce a functional renormalization group equation for the 
one-particle-irreducible effective action, which has a one-loop form.
The master equation is compatible with it,
i.e.~BRST symmetry is preserved along the flow, and it
has a standard regulator-independent Zinn-Justin form.
As a first application, we compute the leading-order gluon wave-function renormalization.
\end{abstract}

\maketitle

\section{Introduction}
\label{sec:intro}

The fact that global symmetries can be implemented exactly is one of the cornerstones of many qualitative and quantitative successes of functional continuum methods in quantum field theory. A prime example is chiral symmetry which can exactly be accounted for even in the presence of ultraviolet (UV) and infrared (IR) regularizations as formulated in the framework of the functional renormalization group (RG)  \cite{Wetterich:1992yh,Bonini:1992vh,Ellwanger:1993mw,Morris:1993qb}. By contrast, local gauge symmetries, as well as nonlinear or diffeomorphism symmetries, require a more careful discussion, as a symmetry transformation can arbitrarily mix modes in momentum space. This seems naturally in conflict with regularizations that operate locally in momentum space. A elementary example is given by a mass term for the gauge boson which would provide for an IR regularization, but breaks gauge invariance.

In standard continuum formulations, quantization of gauge theories, such as Faddeev-Popov quantization, involves a gauge-fixing
procedure in order to remove the large redundancy in the space of field configurations to be integrated over. This goes along with explicit symmetry-breaking terms. While gauge-invariant observables are not affected by the details of the gauge-fixing procedure, gauge-variant building blocks such as gauge-field correlation functions and vertices do depend on the gauge choice. The underlying symmetry is still encoded in Ward-Takahashi identities that relate these correlation functions also across loop orders. While the computation of gauge-invariant observables out of gauge-variant building blocks such as correlation functions thereby remains conceptually possible, it becomes technically more demanding.

A major simplification arises from BRST symmetry, a remnant global supersymmetry that nonlinearly mixes gauge, Faddeev-Popov ghost and further auxiliary fields \cite{Becchi:1974md,Becchi:1975nq,Tyutin:1975qk}. BRST symmetry not only helps identifying the physical Hilbert space of states but also simplifies the constraint equation for correlation functions in the form of the Zinn-Justin master equation~\cite{ZinnJustin:1974mc,ZinnJustin:1975wb}. At the expense of auxiliary sources, the Zinn-Justin equation relates correlation functions algebraically. I.e., its resolution can be approached by algebraic cohomology methods and does not require the computation of loop terms.

In the presence of a generic momentum-space regularization, the elegance (and practicality)  of the master equation is no longer present. For functional RG flows, it has been shown by Ellwanger \cite{Ellwanger:1994iz} that gauge invariance of correlation functions as summarized by the 
effective action can still be encoded in a master equation. However, the regularization procedure which is generically encoded on the level of the propagator leads to additional regulator-dependent terms in the master equation (modified Slavnov-Taylor identities) which again correspond to loop terms. Whereas this modified master equation, as well as a corresponding additional modified Ward-Takahashi identity \cite{Bonini:1993sj,DAttanasio:1996tzp,Litim:1998wk,Litim:1998nf}, encodes the constraints imposed by the symmetries on a conceptually satisfactory level \cite{Pawlowski:2005xe,Gies:2006wv}, it increases the level of technical complexity for nonperturbative approximation schemes, cf. \cite{Ellwanger:1995qf,Ellwanger:1996wy}. Direct applications of functional RG flows together with a resolution of the modified master equation beyond perturbation theory have remained rare \cite{Ellwanger:1995qf,Ellwanger:1996wy,Gies:2003dp,Fischer:2004uk,Fejos:2017sjl}, though the intricacies can simplify in certain gauges such as the Landau gauge \cite{Fischer:2008uz,Cyrol:2016tym}.

Several schemes have been devised, to tackle this practical problem. For instance, an alternative approach has been developed in \cite{Igarashi:1999rm,Sonoda:2007dj,Igarashi:2007fw,Igarashi:2016gcf} introducing a deformed BRST symmetry that includes the regulator and reduces to standard BRST symmetry in the suitable limits. This formulation indeed encodes gauge invariance even of the regularized theory in a bilinear master equation, but at the same time makes the nonlocality of gauge symmetry in momentum space manifest. A convenient scheme to devise nonperturbative approximations relies on the use of background-field methods \cite{Reuter:1993kw}, where invariance under background-field transformations can rather straightforwardly be obtained. Nevertheless, the true quantum gauge invariance is again encoded in modified symmetry identities (Nielsen identities, shift-Ward identity) \cite{Reuter:1994zn,Reuter:1997gx,Freire:2000mn,Freire:2000bq,Pawlowski:2005xe,Bridle:2013sra,Becker:2014qya,Dietz:2015owa,Percacci:2016arh,Safari:2016dwj,Safari:2016gtj}, which in practice have been largely treated on an approximate level \cite{Reuter:1994zn,Reuter:1997gx,Gies:2002af,Braun:2010cy,Eichhorn:2010zc}.

Several further directions have been explored in this context: a manifestly gauge-invariant RG flow has been proposed in \cite{Morris:1998kz} and further developed in \cite{Arnone:2001iy,Arnone:2002cs,Arnone:2005fb} which does not rely on Faddeev-Popov quantization, but makes use of an embedding into an SU($N|N$) supergauge theory. A variety of results has been obtained \cite{Arnone:2003pa,Arnone:2005vd} including a gauge-invariant computation of the two-loop $\beta$ function \cite{Morris:2005tv,Morris:2006in,Evans:2006eq,Rosten:2010vm}. Gauge-invariant RG flows for the geometric effective action have also been set up within the Vilkovisky-DeWitt framework \cite{Branchina:2003ek,Pawlowski:2003sk} with application in the asymptotic-safety scenario for quantum gravity \cite{Donkin:2012ud}. A gauge-invariant RG flow has also recently been constructed in \cite{Wetterich:2016ewc,Wetterich:2017aoy} making use of physical gauges and the freedom to suitably define the macroscopic field and the effective action. Despite these conceptually successful implementations of gauge invariance in RG flows, the most advanced applications to nonperturbative questions often rely on the standard Faddeev-Popov quantization as this has remained technically more accessible for sophisticated systematic expansion schemes, cf. \cite{Cyrol:2016tym,Denz:2016qks,Cyrol:2017ewj,Corell:2018yil}.

In the present work, we suggest a novel approach for the construction of RG flows for gauge systems that relies on Faddeev-Popov quantization and aims at preserving exact BRST symmetry. The key idea is to treat the regularization as a contribution to the gauge fixing. As this is not possible within linear gauges, we consider a special choice of a nonlinear (quadratic) gauge condition. BRST symmetry in the standard fashion remains manifest at all stages of the construction. In order to obtain a conventional regulator term for the gauge field, we use a Fourier noise field for the Nakanishi-Lautrup auxiliary field. As a new ingredient, this gives rise to a background Nakanishi-Lautrup field.

Also the ghost sector is at variance with that of standard linear gauges while preserving exact BRST invariance. In particular, new regulator-dependent vertices appear. Nevertheless, gauge invariance gives rise to a master equation for the (Legendre) effective action which can be brought to standard form with the aid of two additional source fields. This facilitates the resolution of the symmetry constraints by conventional algebraic cohomology methods. Most importantly, the resulting functional nonperturbative flow equation for the effective action has a one-loop structure and is thus amenable to widely used nonperturbative approximation schemes.

The paper is organized as follows. In \Secref{sec:conventions}, we set the stage by recalling basics of gauge-fixed functional integrals in order to introduce our conventions. We introduce the Nakanishi-Lautrup auxiliary field in \Secref{sec:noise} together with a convenient choice of a corresponding noise field which is advantageous for our formalism. \Secref{sec:nonlinearF} is devoted to a discussion of constructing a simple nonlinear gauge that allows to write down an action with BRST-invariant mass terms; since this is a rather widely discussed topic in the literature, this section might be of interest in its own right. Here, we use it as a motivation for the construction of a BRST-invariant RG flow. The latter is presented in \Secref{sec:FRGconstruction}, where we derive the one-loop functional RG flow equation. 
The master equation encoding BRST invariance and its RG flow is
discussed in \Secref{sec:flowmasterequation}.
Here, we give an explicit proof of the compatibility between the master equation and the flow equation, i.e. an action that satisfies the master equation at an initial scale will do so on all scales, provided it also satisfies the flow equation that links the two scales. 
The question concerning the well posedness
of the functional regularization obtained through 
a nonlinear gauge fixing is addressed in \Secref{sec:reconstruction},
where we analyze the mapping between the effective action
and the bare action, and the choice of RG initial 
conditions.
A simple application of the functional flow equation
to one-loop order is presented in \Secref{sec:oneloop}.
Auxiliary information is presented in three
appendices.

\section{Conventions}
\label{sec:conventions}

In this work, we discuss pure Yang-Mills theory in $d$-dimensional
Euclidean spacetime, using the gauge field $A_\mu^a(x)$ in the adjoint
representation of the gauge group $G$ as the local field degree of freedom. For
the counting of canonical dimensions, we implicitly use $d=4$ as an
illustration. Other than that, our formalism applies to general
$d$. The inclusion of charged matter fields is straightforward. The use of the gauge field entails a large redundancy manifested by the invariance under local (infinitesimal) gauge transformations,  $\delta A_\mu^a(x)=\left(D_\mu \omega\right)^a(x)$ for infinitesimal $\omega^a(x)$. Here, we use the covariant derivative in the adjoint representation,
\begin{equation}
D_\mu^{ab}=\partial_\mu\delta^{ab}+g f^{abc}A_\mu^c\, .\label{eq:covder}
\end{equation}
We use condensed notation such that color indices replace also
spacetime indices, and the summation convention over these repeated
indices is extended to integration over the corresponding
spacetime points, whenever two identical indices both refer to field variables. E.g., the covariant derivative then reads
\begin{equation}
D_\mu^{ab}(x_a,x_b)=\left( \delta^{ab}\partial_\mu
+g f^{abc}A_\mu^c(x_a)\right)\delta(x_a-x_b)\, .
\label{eq:covdercondensed}
\end{equation}
Finite gauge transformations can be written as
\begin{equation}
  A_\mu^{\omega}=U A_\mu U^{-1}-\frac{\I}{g}\left(\partial_\mu U\right) U^{-1} \, ,
  \quad A_\mu= A_\mu^a T^a, \label{eq:finiteGT}
\end{equation}
where
\begin{equation}
U(\omega)=e^{-\I g\omega^a T^a}\in G\, , \label{eq:U}
\end{equation}
with general finite $\omega^a(x)$ and generators
\begin{equation}
\left[T^a,T^b\right]=\I f^{abc}T^c\, . \label{eq:LieAlgebra}
\end{equation}
The field strength reads
\begin{equation}
F^a_{\mu\nu}=\partial_\mu A_\nu^a-\partial_\nu A_\mu^a
+g f^{abc}A_\mu^b A_\nu^c, \label{eq:fieldstrength}
\end{equation}
with adjoint indices $a,b,c\dots$. In
condensed notation, the Yang-Mills action is given by
\begin{equation}
S_{\YM}[A]=\frac{1}{4}
F_{\mu\nu}^a F^{a\mu\nu}\, . \label{eq:SYM}
\end{equation}
For quantization, we introduce a gauge-fixing functional $\F^a[A]$, playing a central role for the Faddeev-Popov method:
\begin{equation}
1=\int\!\DD \F^a\,  \delta[\F^a]=\int\!\dd\mu(\omega)\,\delta\left[\F^a[A^\omega]\right]
\Delta_\FP\left[A^\omega\right]\, ,
\label{eq:FaddeevPopov1}
\end{equation}
where $\dd\mu$ is the Haar measure and 
\begin{equation}
  \Delta_\FP\left[A^\omega\right]=\Det\frac{\delta \F^a[A^\omega]}{\delta\omega^b}\, ,
  \label{eq:FPdet}
\end{equation}
is the Faddeev-Popov determinant. The latter is gauge invariant, so that we can replace 
$A^{\omega}$ with $A$.
This determinant can be written in terms of a local action $S_\gh$ by means
of ghost fields
\begin{equation}
\Delta_\FP[A]=\int\!\!\DD\bC\,\DD c\, e^{-S_\gh[A,c,\bC]}\, . \label{eq:FPtoGhosts}
\end{equation}
Imposing a strict gauge-fixing condition is not necessary, because
replacing
\begin{align}
\delta\left[\F^a[A^\omega]\right]&\longrightarrow
B\left[\F^a[A^\omega]\right]
\end{align}
simply changes the $1$ on the left-hand side of \Eqref{eq:FaddeevPopov1} into a
constant. The standard textbook example is
\begin{align}
B\left[\F^a[A]\right]=e^{-S_\gf[A]}\, ,
\end{align}
based on a local gauge-fixing contribution to the action,
\begin{equation}
  S_\gf[A] = \frac{1}{2\xi} \F^a[A]\F^a[A],\label{eq:Sgf}
\end{equation}
with gauge parameter $\xi$. 

Baring explicit breakings through the gauge condition, the global $G$ symmetry remains intact even after
gauge fixing. 
E.g., the ghosts transform under the adjoint of the global $G$ group, i.e. $\omega=\text{const.}$,
\begin{align}
\bs
\delta c^a&= g f^{abc}\omega^b c^c\, ,\\
\delta \bC^a&= g f^{abc}\omega^b \bC^c\, .
\es
\end{align}
We can associate a set of generators to both local gauge and global
color rotations given by
\begin{equation}
\cG^a(x)=\cG^a_A(x)+\cG^a_\gh(x)\, ,\label{eq:gen1}
\end{equation}
where
\begin{align}
\bs
\cG^a_A(x)&=D_\mu^{ab}\frac{\delta}{\delta A_\mu^b},\\
\cG^a_\gh(x)&=-g f^{abc}\left(c^c\frac{\delta }{\delta c^b}+
\bC^c\frac{\delta }{\delta \bC^b} \right) .
\es\label{eq:gen2}
\end{align}
All functional derivatives in this paper are left derivatives by default, unless otherwise specified.
The gauge action $S_\gf+S_\gh$ emerging from 
the Faddeev-Popov construction exhibits an additional global (super-)symmetry: BRST symmetry. Introducing a Grassmannian BRST operator $s$ acting on the fields, the BRST transforms read
\begin{align}
\bs
\left(sA\right)^a_\mu&=D_\mu^{ab}c^b,\\
\left(sc\right)^a&=\frac{1}{2}g f^{abc}c^bc^c,\\
\left(s\bC\right)^a&=- \frac{1}{\xi} \F^a.\\
\es
\end{align}
As the BRST transform of the gauge potential has the form of a gauge transformation, any gauge-invariant contribution to the action is guaranteed to be BRST invariant. 

Quantization now proceeds straightforwardly through the generating functional
\begin{equation}
Z[J,\eta,\bar{\eta}]=e^{W[J,\eta,\bar{\eta}]}=\int\!\DD A \DD c \DD \bC\, e^{-S_\tot}\, ,
\end{equation}
with the total action
\begin{eqnarray}
S_\tot[A,c,\bC,J,\eta,\bar{\eta}]&=&S_\YM[A]+S_\gh[A,c,\bC]+S_\gf[A]
\nonumber\\
& &+S_\so[A,c,\bC,J,\eta,\bar{\eta}]\, .\label{eq:Stot}
\end{eqnarray}
The source terms are summarized in 
\begin{equation}
S_\so[A,c,\bC,J,\eta,\bar{\eta}]=-
\left(
J^{a\mu}A^a_\mu+\bar{\eta}^a c^a+\bC^a\eta^a\right) .
\end{equation}
By Legendre transformation, the effective action can be constructed from the Schwinger functional $W[J,\eta,\bar{\eta}]$,
\begin{equation}
\Gamma[A,c,\bC]=\mathop{\text{sup}}_{J,\eta,\bar{\eta}}\left\{
 J^{a\mu} A_\mu^a +\bar{\eta}^a c^a +\bC^a\eta^a 
 -W[J,\eta,\bar{\eta}]\right\}\!.\quad\quad
 \label{eq:Gamma}
\end{equation}
The effective action is the generating functional for one-particle
irreducible (1PI) proper vertices, being a quantity of central
interest in the following.

\section{Quantization with Fourier noise}
\label{sec:noise}

As the BRST symmetry is a supersymmetry, there is also an ``off-shell'' formulation involving an auxiliary field, the Nakanishi-Lautrup field. The corresponding generalized construction proceeds via the generating functional
\begin{equation}
Z=\!\int\!\!\DD A \DD \C \DD \bC \DD \B
\DD \no \,e^{-S_\YM[A]-S_\text{gauge}[A,\C,\bC ,\B,\no]},\quad
\label{eq:genfuncNL}
\end{equation}
where the generalized gauge-fixing sector is now encoded in
\begin{eqnarray}
S_\text{gauge}[A,\C,\bC,\B,\no ]&=&
S_\gf[A,\B]+ S_\text{noise}[\B,\no]
+S_\gh[\C,\bC],
\quad\quad\label{eq:Sgauge}\\
S_\gf[\B,A]&=&b^a \F^a[A]\,,\label{eq:Sgf2}\\
S_\gh[\C,\bC]&=&-\bC^a {\cal M}^{ab} \C^b
\, ,\quad\quad \label{eq:Sgh2}
\end{eqnarray}
The gauge-fixing action now is linear in the gauge-fixing condition $\F^a$ as well as in the Nakanishi-Lautrup field $\B^a$. We again encounter the Faddeev-Popov operator
\begin{equation}
{\cal M}^{ab}=\left. \frac{\delta \F^a[A]}{\delta A_\mu^c}\frac{\delta A^{\omega\, c}_\mu}{\delta \omega^b}\right|_{\omega=0}
=\frac{\delta \F^a[A]}{\delta A_\mu^c} D_\mu^{cb}\, ,
\end{equation}
and $\no^a$ is a noise field.
We already included in the definition of $Z$ the averaging over the noise, with measure $\exp\{-S_\text{noise}[\B,\no]\}$.
We could equivalently integrate out the noise
and translate this into an action for the
Nakanishi-Lautrup field,
\begin{equation}
e^{-S_\NL[\B]}=\int\!\!\DD \no\, e^{-S_\text{noise}[\B,\no]}\, .\label{eq:SNL}
\end{equation}
Thus, the generating functional reduces to
\begin{equation}
Z=\int\!\!\DD A \DD c\DD\bC \DD b\,
e^{-S[A,c,\bC,b]} ,\label{eq:genfunc3}
\end{equation}
with
\begin{equation}
S[A,c,\bC,b]=S_\YM[A]+S_\gf[A,\B]+S_\NL[\B]+S_\gh[\C,\bC]\,.
\quad\quad
\label{eq:mostgeneralS}
\end{equation}
A Gaussian weight for the noise
\begin{equation}
S_\text{noise}[\B,\no]=\frac{1}{2\xi}\no^a\no^a-\I\B^a\no^a\, ,
\label{eq:Gaussianweight}
\end{equation}
corresponds to a local action for the Nakanishi-Lautrup field 
\begin{equation}
S_\NL[\B]=\frac{\xi}{2}\B^a \B^a\, ,\label{eq:SNL2}
\end{equation}
highlighting its auxiliary-field character. Upon integrating out the $\B$ field, this entails
\begin{equation}
S_\gf[A]=\frac{1}{2\xi}\F^a[A]\F^a[A]\,, \label{eq:SF2}
\end{equation}
demonstrating the equivalence to the preceding section. 
In the present work, we focus instead 
on the choice 
\begin{equation}
S_\text{noise}[\B,\no]=\I\left(v^a-\B^a\right)\no^a\, ,
\label{eq:Fourierweight}
\end{equation}
where $v^a$ is an external vector field. 
This leads to
a Fourier weight that
results in
\begin{equation}
e^{-S_\NL[\B]}=\delta\!\left[\B^a-v^a\right]\, ,\label{eq:SNL3}
\end{equation}
which, after integration of $\B$, translates into
\begin{equation}
S_\gf[A]=v^a\F^a[A]\,.
\label{eq:Sgflinear}
\end{equation}
We observe that $S_\gf$ remains linear at the expense of introducing an external field $v^a$. Even though we are interested in a nonlinear gauge-fixing
functional $\F^a$, we choose conventions such that $\F^a$ retains its standard canonical dimension $[\F^a]=2$, implying a corresponding dimension $[v]=2$.
The vector $v^a$ can be interpreted as an external field,
which explicitly breaks the global $G$ symmetry. 

For any $S_\NL$, the action of \Eqref{eq:mostgeneralS}
is invariant under the following BRST symmetry
\begin{equation}
\begin{aligned}
sA^a_\mu&=D_\mu^{ab}c^b\,,&\quad\quad
sc^a&=\frac{g}{2}f^{abc}c^bc^c\, ,\\
sb^a&=0\,,&\quad\quad
s\bC^a&=b^a\, .
\end{aligned}
\label{eq:brst}
\end{equation} 
The BRST operator is nilpotent, i.e.~$s^2=0$, thanks to the algebraic property
\begin{equation}
\frac{\delta D_\mu^{ab}}{\delta A_\nu^c}D_\nu^{cd}-
\frac{\delta D_\mu^{ad}}{\delta A_\nu^c}D_\nu^{cb}=
g f^{cbd}D_\mu^{ac}\, .
\label{eq:algebraofD}
\end{equation}
Alternatively, it is useful to formulate the symmetry transformation with the 
help of an off-shell BRST generator:
\begin{equation}
\DD_0=\left(D_\mu c\right)^a\frac{\delta\ \ }{\delta A^a_\mu}
+\frac{g}{2}f^{abc}c^bc^c\frac{\delta\ \, }{\delta c^a}
+b^a \frac{\delta\ \, }{\delta \bC^a}\, ,
\label{eq:DD0}
\end{equation}
which is also nilpontent $\DD_0^2=0$. Contrary to standard off-shell
supersymmetry transformations, the BRST symmetry is not a linear
symmetry operation on the fields. This is the main source of
nonlocalities arising in momentum space. For an approach to a linear
version of BRST symmetry, see \cite{Tissier:2009sm}.

If one chooses a Fourier weight for $\no^a$,
as in \Eqref{eq:Fourierweight}, integrating out $\B$
leads to an on-shell action $S[A,c,\bC,v]$
and its corresponding on-shell BRST transformation,
which is obtained from \Eqref{eq:brst} and 
\Eqref{eq:DD0} upon replacement of
$\B^a$ with $v^a$.
Thus, again this BRST transformation is nilpotent, as $sv^a=0$.

Following \cite{ZinnJustin:1974mc,ZinnJustin:1975wb}, we add sources for
both the elementary fields and their BRST variations to $S[A,c,\bC,v]$.
Defining
\begin{align}
\bs
S[A,\C,\bC,v,K,L]&=S[A,\C,\bC,v]\\
&+K_\mu^a\left(D^\mu c\right)^a
+L^a \frac{g}{2} f^{abc} c^b c^c\, ,
\label{eq:mostgenSplusKL}
\es
\end{align}
with $K_\mu^a$ being Grassmann-valued, we now obtain the source part of the action
\begin{align}
\bs
S_\so=&
-J^\mu_a A^a_\mu-\bar{\eta}^a c^a-\bC^a\eta^a\\
&+K_\mu^a\left(D^\mu c\right)^a
+L^a \frac{1}{2}g f^{abc} c^b c^c
\, .
\es
\label{eq:source2}
\end{align}
The generating functional then reads
\begin{equation}
e^{W[J,\eta,\bar{\eta},v,K,L]}=\int\!\!\DD A\DD c\DD\bC\,
e^{-S[A,c,\bC,v]-S_\so}\, .
\label{eq:genfunc4}
\end{equation}
To deduce the master equation, i.e.~the Ward identity for
BRST symmetry,
we change variables of integration according to an infinitesimal
BRST transform. Based on BRST invariance of the measure, we obtain
\begin{equation}
J^a_\mu\langle\left(D^\mu c\right)^a\rangle
-\bar{\eta}^a\langle\frac{g}{2}f^{abc}c^bc^c\rangle
+ v^a \eta^a=0\, .
\label{eq:BRSTvar}
\end{equation}
The sign of the second term in the last equation
comes from  commuting the BRST operator $s$
(or a corresponding Grassmann parameter, say $\bth$)
with $\bar{\eta}$.
In terms of the Schwinger functional, we get
\begin{equation}
-J^a_\mu\frac{\delta W}{\delta K^a_\mu}
+\bar{\eta}^a \frac{\delta W}{\delta L^a}
+v^a \eta^a=0\, .
\label{eq:BRSTvarW}
\end{equation}
Now let us define the effective action
\begin{align}
\begin{split}
\Gamma[A,c,\bC,b,K,L]=\mathop{\text{sup}}_{J,\eta,\bar{\eta}}&
\Big\{
J^\mu_a A^a_\mu+
\bar{\eta}^a c^a+\bC^a\eta^a
\\
&-W[J,\eta,\bar{\eta},K,L]
\Big\}\,,
\end{split}
\label{eq:effact}
\end{align}
such that the ``macroscopic'' fields conjugate to the sources satisfy
\begin{align}
A^\mu_a=\frac{\delta W}{\delta J^a_\mu}\, ,\quad
c^a=\frac{\delta W}{\delta \bar{\eta}^a}\, ,\quad
\bC^a=-\frac{\delta W}{\delta {\eta}^a}= W\mathop{\frac{\delta\ }{\delta {\eta}^a}}^{\leftarrow}\, .
\label{eq:macrofields}
\end{align}
This implies the quantum equations of motion in terms of the 
effective action
\begin{align}
J^a_\mu=\frac{\delta \Gamma}{\delta A^\mu_a}\, ,\quad
{\eta}^a=\frac{\delta \Gamma}{\delta \bC^a}\, ,\quad
\bar{\eta}^a=-\frac{\delta \Gamma}{\delta c^a}=\Gamma\mathop{\frac{\delta\ }{\delta c^a}}^{\leftarrow}\, ,
\label{eq:QoM}
\end{align}
and also the relations
\begin{equation}
\frac{\delta \Gamma}{\delta K^a_\mu}=-\frac{\delta W}{\delta K^a_\mu}\,,
\quad\quad\quad
\frac{\delta \Gamma}{\delta L^a}=-\frac{\delta W}{\delta L^a}\,.
\label{eq:WtoGamma}
\end{equation}
Thus, the Zinn-Justin master equation following with these identities
from \Eqref{eq:BRSTvarW} reads
\begin{equation}
\frac{\delta \Gamma}{\delta A^\mu_a}\frac{\delta \Gamma}{\delta K^a_\mu}
+\frac{\delta \Gamma}{\delta c^a}\frac{\delta \Gamma}{\delta L^a}
+v^a \frac{\delta \Gamma}{\delta \bC^a}=0\,.
\label{eq:mastereqGamma}
\end{equation}
Notice that the BRST invariance of $S[A,c,\bC,b]$,
defined in \Eqref{eq:mostgeneralS}, is encoded in the
following identity fulfilled by 
$S[A,c,\bC,v,K,L]$ of \Eqref{eq:mostgenSplusKL}
\begin{equation}
\frac{\delta S}{\delta A^\mu_a}\frac{\delta S}{\delta K^a_\mu}
+\frac{\delta S}{\delta c^a}\frac{\delta S}{\delta L^a}
+v^a \frac{\delta S}{\delta \bC^a}=0\,.
\label{eq:mastereqS}
\end{equation}
Therefore $\Gamma=S[A,c,\bC,v,K,L]$ is a special solution 
of the master equation~(\ref{eq:mastereqGamma}).

\section{Mass and nonlinear gauge fixing}
\label{sec:nonlinearF}

In the following, we suggest to introduce mass terms by means of the gauge-fixing sector. The problem of constructing a BRST-invariant RG flow is closely related to that of a BRST-invariant mass, since a mass-like regularization is in correspondence to a Callan-Symanzik flow. Our basic idea can thus already be understood on the level of mass terms for the gluon and ghost fields. In fact, such mass terms and their (in-)compatibility with BRST symmetry has been widely discussed in the literature \cite{Curci:1976bt,Curci:1976kh,Ojima:1981fs,Delbourgo:1987np,Blasi:1995vt,Alkofer:2000wg,Kondo:2001nq,Ellwanger:2002sj,Fischer:2008uz,Aguilar:2011ux,Maas:2011se,Aguilar:2015bud}. Recently, the idea has been investigated extensively that the nonperturbative generation of such masses in the propagators could effectively cure the shortcomings of perturbative Faddeev-Popov quantization, most prominently those arising from the Gribov ambiguity. In fact, results from simple perturbation theory based on massive propagators compares rather favorably with lattice simulations \cite{Cucchieri:2007md,Bogolubsky:2009dc,Tissier:2010ts,Boucaud:2011ug,Cucchieri:2011ig,Tissier:2011ey,Pelaez:2014mxa,Reinosa:2013twa,Reinosa:2014ooa,Reinosa:2017qtf}. 

Based on the Fourier weight for the noise of \Eqref{eq:Fourierweight},
which results in the gauge-fixing action in \Eqref{eq:Sgflinear},
it is suggestive to accommodate a mass-like term in a nonlinear gauge fixing
\begin{equation}
  \label{eq:genericF}
\F^a[A] =A^{b\mu} \X^{abc}_{\mu\nu}A^{c\nu} +\LL^{ab}_\mu A^{a\mu} 
\end{equation}
We assume that the matrix $\X^{abc}_{\mu\nu}$ does not 
 add another explicit breaking of the global $G$ symmetry beyond
the one already introduced by $v^a$.
Thus, we assume that it can be written in terms of the $v^a$ vector itself.
As far as the Lorentz symmetry is concerned, non-covariant gauges 
can be easily embedded into this ansatz, for instance by choosing either $\LL$ or
$\X$ or both to depend on a specific spacetime vector.
Yet, in this work we choose to discuss examples corresponding to covariant
gauges where this breaking does not occur.
Furthermore, we choose $\X^{abc}_{\mu\nu}$ to
be  always proportional to $\delta(x_a-x_b)\delta(x_b-x_c)$,
such that $\F^a$ is a local functional,
which depends on $A$ and $v$ at the spacetime
 point $x_a$ only.
To simplify notations in the following we always drop
these delta functions.

The choice on which we focus in this section to illustrate the properties of the
construction is 
\begin{align}
\bs
\X^{abc}_{\mu\nu}&=\frac{v^a }{2|v|^2}
\left(
\m^2\delta_{\mu\nu}-\frac{1}{\xi}\partial_\mu\partial_\nu
\right)\delta^{bc}\,,\\
\LL^{ab}_\mu&=\left(1+\frac{\mm^2}{-\partial^2}\right)\partial_\mu\delta^{ab}\,.
\es
\label{eq:Fwithmasses}
\end{align}
This particular example leads
to a gluonic sector of the bare action which is simply a Yang-Mills action
with a Lorenz gauge fixing plus a mass-like parameter for both the 
longitudinal and the transverse vector bosons
\begin{align}
\bs
S_\gf[A]=&\frac{1}{2}\m^2 A^a_\mu A^{a\mu}+\frac{1}{2\xi}\left(\partial^\mu\! A^{a}_\mu\right)^2\\
&+v^a\Big(1+\frac{\mm^2}{-\partial^2}\Big)\partial^\mu\! A^{a}_\mu\,.
\es
\label{eq:Sgfmasslike}
\end{align}
The last term in this expression is an unusual linear
shift of the action which 
is added to the action
of a vector field with mass $\m$,
\begin{align}
\bs
&\left(S_\YM+S_\gf\right)[A]=v^a\Big(1+\frac{\mm^2}{-\partial^2}\Big)\partial^\mu\! A^{a}_{\mu}\\
&+\frac{1}{2} A^a_\mu\left(-\partial^2+\m^2\right)
\big(\Pi_{\T}+\frac{1}{\xi}\Pi_{\Lo}\big)^{\mu\nu} A^{a}_\nu+O(A^3)
\es
\label{eq:Sproca}
\end{align}
where 
\begin{equation}
\Pi_{\Lo}^{\mu\nu}=\frac{\partial^\mu \partial^\nu}{\partial^2}\,,\qquad \Pi_{\T}^{\mu\nu}=\delta^{\mu\nu}-\frac{\partial^\mu \partial^\nu}{\partial^2}\,.
\label{eq:projectors}
\end{equation}
The linear shift in \Eqref{eq:Sproca} plays the role of an external source
\begin{equation}
  J_\mu^a=\partial_\mu\Big(1+\frac{\mm^2}{-\partial^2}\Big) v^a\,.
  \label{eq:sourceJ}
\end{equation}
In particular, if we set the gluon mass $\m=0$ and
take the $\xi\to+\infty$ limit, the 
classical equation of motion becomes
\begin{equation}
D_\nu^{ab} F^{b\nu\mu}
=J^{a\mu}\,,
\end{equation}
and the classical gauge symmetry
requires 
\begin{equation}
D_\mu^{ab} J^{b\mu}=D_\mu^{ab} \partial^\mu
\Big(1+\frac{\mm^2}{-\partial^2}\Big)  v^b=0.
\label{eq:continuity}
\end{equation}
on the equations of motion.
Thus, even if the source term breaks
the global color symmetry of the action explicitly, charge conservation
is preserved if $v^a$ is chosen to fulfill
\Eqref{eq:continuity}.
At $A=0$, or in the Abelian theory, this requires
\begin{equation}
  (-\partial^2 +\mm^2)v^a=0\,,
  \label{eq:vKG}
\end{equation}
which is a massive Klein-Gordon equation for each component of $v^a$.

While a nonvanishing $v^a$ is a source of explicit symmetry breaking
of global color rotations, the action remains of course form-invariant under global $G$ transformations that include rotations of the $v^a$ field:
%
%
\begin{align}
\bs
\delta_\omega A_\mu^a &= -gf^{abc}A_\mu^b\omega^c\\
\delta_\omega \bar{c}^a &= -gf^{abc}\bar{c}^b\omega^c\\
\delta_\omega c^a &= -gf^{abc}c^b\omega^c\\
\delta_\omega v^a &= -gf^{abc}v^b\omega^c
\es
\end{align}
with $\omega$ a constant infinitesimal parameter.  This should be
contrasted to the behavior under the BRST symmetry, which remains a
symmetry of the action even in the presence of a nonvanishing $v^a$,
simply because it does not require any change of $v^a$.

The ghost action corresponding to \Eqref{eq:Sgfmasslike}
reads 
\begin{align}
\bs
&S_\gh=-\bC^a\Big(1+\frac{\mm^2}{-\partial^2}\Big)\partial^\mu \left(D_\mu c\right)^a\\
&-\frac{v^a }{|v|^2}\bC^a \left(\m^2 A^{b\mu}+\frac{1}{\xi}\left(\partial^\nu\! A^{b}_\nu\right) \partial^\mu \right)\left(D_\mu c\right)^b\,,
\es
\end{align}
revealing $\mm$ to be a ghost mass parameter. Integrating the derivative term of $\F^a$  by parts, the action assumes an alternative form,
\begin{align}
&S_\gh=-\bC^a\Big(1+\frac{\mm^2}{-\partial^2}\Big)\partial^\mu \left(D_\mu c\right)^a\\
&-\frac{v^a }{2|v|^2}\bC^a \!\left[2\m^2 A^{b\mu}
+\frac{1}{\xi} \left(A^{b}_\nu \partial^\nu\partial^\mu+
\partial^\nu\partial^\mu A^{b}_\nu\right)
\right]\!
\left(D_\mu c\right)^b\,.
\nonumber
\end{align}
In both forms the actions contain higher-derivative
interaction terms which are accompanied by the external field $v^a$ carrying a
new scale because of its canonical dimension. 
Furthermore the ghost-mass parameter $\mm^2$
introduces a nonlocal modification of the ghost-gluon vertex. This can be reinterpreted as a deformation of 
the conventional Feynman rules in the Lorenz gauge
according to the following replacement
for the momentum of the antighost
\begin{equation}
  p_\mu\longrightarrow \Big(1+\frac{\mm^2}{p^2}\Big)p_\mu\,.
  \label{eq:momrepl}
\end{equation}
Thus higher-derivative interactions, a new external field,  and nonlocal
vertices are the prices
to be payed for introducing a mass term and thus an IR regularization of vector
and ghost propagators in our approach while preserving BRST symmetry.

It should be noted at this point, that there is another
independent possibility for the introduction 
of a mass parameter for the ghosts.
While \Eqref{eq:Fwithmasses} defines $\bar{m}_\gh$
as part of the gauge-fixing functional,
one might simply add a massive
deformation of $S_\gh$
at fixed $S_\gf$. The problem in this case is
preserving BRST symmetry.
This can be achieved in presence of the background
field $v^a$.
As an example, we observe that the deformation
\begin{equation}
S_\gh\mapsto S_\gh + \left(\bar{m}_\gh' v^a f^{abc}c^b c^c
-\bar{m}_\gh'{\!}^\ast\, v^a f^{abc} \bC^b \bC^c\right)
\end{equation}
is BRST exact.  After diagonalization, this term contributes a
positive mass to the propagator of those ghost fields with adjoint
colors perpendicular to $v$. Further BRST-closed ghost bilinears
involving also a background field $v^a$ may be conceivable, but will be
discussed elsewhere.

As we have introduced the mass terms through the gauge condition, it is instructive to take a second look at the gauge condition in terms of transverse and longitudinal components:
\begin{align}
v^a\F^a=&\frac{1}{2}\m^2 A^a_{\T\mu} A^{a\mu}_{\T}+\frac{1}{2}\m^2 A^a_{\Lo\mu} A^{a\mu}_{\Lo}+\frac{1}{2\xi}\left(\partial^\mu\! A^{a}_{\Lo\mu}\right)^2
\nonumber\\
&+v^a\Big(1+\frac{\mm^2}{-\partial^2}\Big)\partial^\mu\! A^{a}_{\Lo\mu}\,.
\label{eq:SgfmasslikeTL}
\end{align}
In order to satisfy the gauge condition $\F^a=0$, a field
configuration must fulfill $v^a\F^a=0$. The terms in the first line of
\Eqref{eq:SgfmasslikeTL} are manifestly positive, whereas the last
term can have either sign. While for vanishing mass terms, the gauge
condition $\F^a=0$ would correspond to $\partial_\mu A_{\Lo}^{a\mu}=0$
as usual, the finite mass version requires a cancellation of the first
line against the second line. For any finite gauge parameter and any
transversal field content $A_{\T}$, it is conceivable that a
longitudinal field content $A_{\Lo}$ can be gauged
accordingly. However, in the Landau gauge limit $\xi\to0$, the
third term strictly enforces $\partial_\mu A_{\Lo}^{a\mu}=0$, making
it impossible to satisfy the gauge condition for $\bar{m}\neq 0$. 
While we have introduced the mass terms through the gauge
condition, this argument demonstrates that this condition does not
strictly speaking define an ordinary gauge. The standard 
Landau-gauge-fixed functional integral is only recovered in the limit
of vanishing masses. Nevertheless, BRST symmetry is preserved at all
stages of the construction.

\section{The functional renormalization group}
\label{sec:FRGconstruction}

\subsection{Regularization and nonlinear gauge fixing}
\label{sec:regularization}

Let us now address simultaneously the task of gauge fixing and
regularizing the functional integral in a BRST-symmetric manner.
%
Based on the Fourier weight for the noise of \Eqref{eq:Fourierweight},
which results in the gauge-fixing action in \Eqref{eq:Sgflinear},
we still adopt a gauge-fixing functional of the 
form in \Eqref{eq:genericF}.
To regularize not only IR but also UV divergences we
take inspiration from the massive gauge fixing
of \Eqref{eq:Fwithmasses}, but now 
we promote the mass parameters to arbitrary
derivative kernels:
$\bar{m}^2\delta_{\mu\nu}\to R_{\mu\nu}(\partial)$
and $\bar{m}_\gh^2\to R_\gh(\partial)=(-\partial^2)r_\gh(-\partial^2)$.
In other words, we suggest a gauge condition of the following form:
\begin{align}
\bs
\X^{abc}_{\mu\nu}&=\frac{v^a }{2|v|^2}\X_{\mu\nu}
\delta^{bc}\,,\\
\LL^{ab}_\mu&=\left(1+r_\gh(-\partial^2)\right)\partial_\mu\delta^{ab}\,.
\es
\label{eq:frgQL}
\end{align}
where
\begin{equation}
\X_{\mu\nu}=
R_{\mu\nu}(\partial)-\frac{1}{\xi}\partial_\mu\partial_\nu\,.
\label{eq:frgX}
\end{equation}
Here $R_{\mu\nu}$ is a symmetric tensor and an even differential operator.
%
%
%
To be more specific, a possible choice
for it reads
\begin{equation}
R^{\mu\nu}(\partial)=
R_{\Lo}(-\partial^2) \Pi_{\Lo}^{\mu\nu} +
R_{\T}(-\partial^2)\Pi_{\T}^{\mu\nu}.
\label{eq:Rmunu}
\end{equation}
The functions $R_{\gh,\T,\Lo}$ are
regulators in momentum space known from the construction of the
Wetterich equation \cite{Wetterich:1992yh}, being an RG flow equation
for an effective action depending on a regulator scale $k$. The
regulators implement the momentum space regularization by providing a
mass gap to modes with momenta $p^2\lesssim k^2$, by satisfying
$\lim_{p^2/k^2\to 0} R_{\gh,\T,\Lo}(p^2) > 0$. By contrast, the
momentum modes beyond the RG scale $k$ should remain essentially
unmodified, $\lim_{k^2/p^2\to 0} R_{\gh,\T,\Lo}(p^2) = 0$. For the
scale $k$ approaching the UV regularization scale $k\to \Lambda$
(possibly with the limit $\Lambda\to\infty$), the regulator function
should diverge, thereby suppressing all quantum fluctuations such that
the effective action can be matched with the classical action to be
quantized; for reviews, see \cite{Berges:2000ew,Pawlowski:2005xe,Gies:2006wv,Delamotte:2007pf,Braun:2011pp,Nagy:2012ef}. 
More details on the UV limit are discussed in \Secref{sec:reconstruction}.

The gauge-fixing and ghost actions corresponding
to this choice of gauge-fixing functional read
\begin{align}
&S_{\gf}= \frac{1}{2}{A}_\mu^a \X^{\mu\nu}\! A^{a}_\nu + v^a (1+r_\gh(-\partial^2)) \partial^\mu\! A^{a}_\mu,\label{eq:SgfFRG}\\
\bs
&S_{\gh}= -\bar{c}^a (1+r_\gh(-\partial^2)) \left(\partial^\mu D_\mu c\right)^a 
\label{eq:SghFRG}\\
&-\frac{v^a}{2|v|^2}\bar{c}^a \left(
\left(\X^{\mu\nu}{A}^b_\nu\right) \left({D}_{\mu}c\right)^b
+{A}^b_\mu\left(\X^{\mu\nu}{D}_{\nu}c\right)^b 
\right).
\es
\end{align}
In contrast to the standard construction of flow equations, the regulators now appear also in ghost-gluon vertex operators.
In order to keep the subsequent flow equation at most on the two-point level, 
we need two extra sources in the partition function. For the on-shell case, we thus work with the generating functional:
\begin{equation}
e^{W[J,\eta,\bar{\eta},v,K,L,M,I]}=\int\!\DD A\DD c\DD\bC\,
e^{-S[A,c,\bC,v]-S_\so} \, .
\label{eq:frgpartition}
\end{equation}
where the source terms now read
\begin{align}
\bs
S_\so=&-J^{a \mu} A^a_\mu-\bar{\eta}^a c^a-\bC^a\eta^a\\
&+K_\mu^a\left(D^\mu c\right)^a
+L^a \frac{g}{2} f^{abc} c^b c^c\\
&-\frac{v^a}{|v|^2}\bar{c}^a{A}^b_\mu I^{b\mu} 
-\frac{v^a}{|v|^2} \bar{c}^a \left({D}_\mu c\right)^b M^{b\mu}\, .
\label{eq:Ssoexplicit}
\es
\end{align}
and the action includes the ghost and gauge-fixing parts displayed in Eqs.~\eqref{eq:SgfFRG} and \eqref{eq:SghFRG}:
\begin{equation}
S[A,\bar{c},c,v]=S_{\YM}[A]+S_{\gf}[A,v]+S_{\gh}[A,c,\bC,v]\,.
\label{eq:bareS}
\end{equation}
Notice that $I^b_\mu$ and $K^a_\mu$ are anticommuting,
while $L^a$ and $M^b_\mu$ are commuting sources.
Recall that $S[A,c,\bC,v]$ is invariant 
under the BRST transformation
\begin{align}
\bs
s A^a_\mu &=  D_\mu c^a\,,\\
s c^a &= \frac{g}{2} f^{abc} c^b c^c\,,\\
s \bC^a &=  v^a\,, \\
s v^a &= 0\,.
\es
\label{eq:BRSTagain}
\end{align}
To highlight the BRST properties of the 
composite operators introduced in \Eqref{eq:Ssoexplicit},
we denote
%
%
%
\begin{align}
\Omega^b_\mu&=\frac{v^a}{|v|^2}\bar{c}^a{A}^b_\mu\,,\\
\mathcal{A}^{b}_\mu&=\frac{v^a}{|v|^2} \bar{c}^a \left({D}_\mu c\right)^b\,.
\end{align}
and define
\begin{align}
\bs
S_\so^\brs= 
K_\mu^a \, s A^a_\mu
+L^a \, s c^a
-\Omega^b_\mu  I^{b\mu} 
- \mathcal{A}^{b}_\mu M^{b\mu}\, .
\label{eq:Ssobrst}
\es
\end{align}
To determine the BRST variation of this additional source action
we notice the following peculiar structure:
\begin{align}
s \Omega^a_\mu&=A^a_\mu-\mathcal{A}^{a}_\mu \,,\\
s \mathcal{A}^{a}_\mu&=s A^{a}_\mu\,.
\end{align}
Thus $A$ and $\mathcal{A}$ 
are cohomologous to each other, as their difference
is BRST exact. In other words, they belong to the 
same BRST cohomology class.

The encoding of the BRST symmetry of the bare action
in the properties of the effective action and of its
RG flow will be discussed in \Secref{sec:flowmasterequation}.
Before that, let us deduce from the regularization
we have provided, an exact RG equation for
the effective action.

\subsection{Flow of the effective average action}
\label{sec:EAA}

To write the flow equation for the 1PI effective action it is 
useful to collect the sources into vectors
\be
\cJ^{\dagger}_i=\left(J^{a}_\mu,\bar{\eta}^a,\eta^a\right)\,,
\quad\quad\quad
\cJ_i=\begin{pmatrix}
	J^{a}_\mu\\ 
	\bar{\eta}^a\\
	\eta^a
\end{pmatrix},\quad
\ee
and correspondingly
define collective fields
\be
\Phi^{\dagger i}=\left(A^{a\mu},-c^a,\bC^a\right)\,,
\quad\quad\quad
\Phi^i=\begin{pmatrix}
	A^{a\mu}\\ 
	c^a\\
	-\bC^a
\end{pmatrix}.\quad
\ee
It is also convenient to group the sources for composite operators into vectors
\be
\cI^{\dagger}_i=\left(K^{a}_\mu,L^a, M^a_\mu, I^a_\mu\right)\,,
\quad\quad\quad
\cI_i=\begin{pmatrix}
	K^{a}_\mu\\ 
	L^a\\
	M^a_\mu\\
	I^a_\mu
\end{pmatrix}. \quad
\label{eq:compsource}
\ee
Here and in the following the latin letters $i,j,k,\dots$
are adopted for collective indices which refer to the components of
$\cJ,\cI,\Phi$ as well as to their spacetime and/or color indices,
and to the spacetime point at which they are evaluated.
From now on, we follow the common convention used in functional renormalization to denote the Legendre transform of the regularized Schwinger functional $W$
as $\tGamma$, while $\Gamma$ is reserved for 
$\tGamma$ minus the regulator terms.
In formulas, we define the Legendre effective action as
%
%
%
\begin{align}
\label{eq:deftGamma}
\bs
\tGamma[\Phi,v,\cI]=\mathop{\text{sup}}_{\cJ_i}\left\{
\cJ_i^\dagger\Phi^i -W[\cJ,v,\cI]
\right\}\,,
\es
\end{align}
and the effective average action \cite{Wetterich:1992yh} as
\begin{align}
\label{eq:defGamma}
\bs
\Gamma[\Phi,v,\cI]=\,\tGamma[\Phi,v,\cI]-\Delta S[\Phi,v]\,.
\es
\end{align}
In the second definition, we have introduced an abbreviation for the regulator contribution to the action
\begin{equation}
\Delta S=\Delta S_\gf+\Delta S_\gh\, ,
\label{eq:defDeltaS}
\end{equation}
where the scale dependence in the functional integral arises from the regulators
\begin{align}
&\Delta S_{\gf}= \frac{1}{2}{A}_\mu^a R^{\mu\nu}\! A^{a}_\nu +  v^a r_\gh(-\partial^2) \partial^\mu\! A^{a}_\mu\, ,
\label{eq:DeltaSgf}\\
\bs
&\Delta S_{\gh}= -\bar{c}^a r_\gh(-\partial^2) \left(\partial^\mu D_\mu c\right)^a \\
&-\frac{v^a}{2|v|^2}\bar{c}^a \left(
\left(R^{\mu\nu}{A}^b_\nu\right) \left({D}_{\mu}c\right)^b
+{A}^b_\mu\left(R^{\mu\nu}{D}_{\nu}c\right)^b 
\right).
\label{eq:DeltaSgh}
\es
\end{align}
We emphasize that the Legendre transform in Eqs.~\eqref{eq:deftGamma} and \eqref{eq:defGamma} is just performed with respect to the sources for the elementary gauge and ghost fields. In this way, our action remains on the 1PI level.

The Legendre transform of \Eqref{eq:deftGamma} 
translates into standard formulas connecting
derivatives of $W$ to derivatives of $\tGamma$,
such as
\begin{align}
\frac{\delta\ }{\delta \cJ_i^\dagger}W&=\Phi^i\,,\quad\quad
W \mathop{\frac{\delta\ }{\delta \cJ_i}}^{\leftarrow}=\Phi^{\dagger i}\,,
\label{eq:W1=Phi}\\
\frac{\delta\ }{\delta \Phi^{\dagger i}}\tGamma&=\cJ_i\,,\quad\quad
\tGamma \mathop{\frac{\delta\ }{\delta \Phi^i}}^{\leftarrow}=\cJ^{\dagger}_i\,.
\label{eq:Gamma1=J}
\end{align}
Then we can denote
\begin{align}
W^{(2)}_{\cJ_i \cJ_j}
&=\frac{\delta\ }{\delta \cJ_i^\dagger}W\mathop{\frac{\delta\ }{\delta \cJ_j}}^{\leftarrow}\,,
\label{eq:W2}\\
\tGamma^{(2)}_{\Phi^i \Phi^j}
&=\frac{\delta\ }{\delta \Phi^{\dagger i}}\tGamma
\mathop{\frac{\delta\ }{\delta \Phi^j}}^{\leftarrow}\,.
\label{eq:Gamma2}
\end{align}
By differentiating \Eqref{eq:W1=Phi} w.r.t. $\Phi$, one deduces
\begin{equation}
W^{(2)}_{\cJ_i \cJ_j}\tGamma^{(2)}_{\Phi^j \Phi^k}=\delta^i_{\ k}\, ,\quad\quad\quad
\tGamma^{(2)}_{\Phi^i \Phi^j}W^{(2)}_{\cJ_j \cJ_k}=\delta^{\ k}_{i}\, ,
\end{equation}
or in other words
\begin{equation}
W^{(2)}_{J_i J_j}=\left(\tGamma^{(2)-1}_{\Phi \Phi}\right)_{i j}
=\left(\tGamma^{(2)-1}\right)_{\Phi^i \Phi^j}\, ,
\end{equation}
where the second equal sign is nothing but a definition of a convenient notation used below.

While in position space one can think of the condensed indices
in these formulas as corresponding to a given coordinate $x$,
more care is needed in momentum space.
There, it is convenient to accompany the $\dagger$ operation
with a reflection of the Fourier momentum.
More details about the Fourier conventions and the 
representation of the relevant formulas in momentum space
are provided in \Appref{app:Fourier}.

As for the sources of \Eqref{eq:compsource}, we have analogously to \Eqref{eq:WtoGamma}
\begin{align}
W^{(2)}_{\cI_i \cI_j}=\frac{\delta\ }{\delta \cI_i^\dagger}W\mathop{\frac{\delta\ }{\delta \cI_j}}^{\leftarrow}
=-\frac{\delta\ }{\delta \cI^{\dagger i}}\tGamma
\mathop{\frac{\delta\ }{\delta \cI^j}}^{\leftarrow}
=-\tGamma^{(2)}_{\cI_i \cI_j}\,.
\label{eq:W2II}
\end{align}
To construct the full matrix of second-order derivatives of $W$,
in terms of second-order derivatives of $\tGamma$, we need expressions
for mixed derivatives.
These descend from differentiating \Eqref{eq:W1=Phi} with respect to $\cI_j$
and accounting for \Eqref{eq:WtoGamma}, yielding
\begin{align}
W^{(2)}_{\cJ_i \cI_k}=&-\big(\tGamma^{(2)-1}_{\Phi \Phi}\big)_{i j}
\tGamma^{(2)}_{\Phi^j \cI_k}\, ,\label{eq:W2JI}\\
W^{(2)}_{\cI_i \cJ_k}=&-\tGamma^{(2)}_{\cI_i \Phi^j}
\big(\tGamma^{(2)-1}_{\Phi \Phi}\big)_{jk}
\, .\label{eq:W2IJ}
\end{align}
Based on these identities, the RG flow equation for $\tGamma$ follows analogously to the standard derivation \cite{Wetterich:1992yh,Bonini:1992vh,Ellwanger:1993mw,Berges:2000ew,Pawlowski:2005xe,Gies:2006wv,Delamotte:2007pf,Braun:2011pp,Nagy:2012ef}, as briefly sketched for our purposes in the following: starting from the partition function \Eqref{eq:frgpartition},
\begin{equation}
Z[\cJ,v,\cI]=e^{W[\cJ,v,\cI]}\, ,
\label{eq:Zoncemore}
\end{equation}
we deduce the flow of $Z$ by differentiating \Eqref{eq:frgpartition} with respect to  $t=\log k$,
\begin{equation}
\partial_t Z=\G_t Z\, .
\label{eq:flowZ}
\end{equation}
Here, we have introduced the generator of RG transformations acting on the partition
function,
\begin{align}
&\G_t=
\left(\partial_t r_\gh\partial_\mu\delta^{ab}\right) 
\left(v^b\frac{\delta\phantom{Z}}{\delta J^a_\mu}
-\frac{\delta^2\phantom{ZZ}}{\delta \eta^{b}\delta K^{a}_\mu}
\right)
\label{eq:GonZ}
\\
&-\frac{1}{2}\left(\partial_t R_{\mu\nu}\delta^{ab}\right)
\left(
\frac{\delta^2\phantom{ZZ}}{\delta J^{a}_\mu\delta J^{b}_\nu}
-\frac{\delta^2\phantom{ZZ}}{\delta M^{a}_\mu\delta J^{b}_\nu}
-\frac{\delta^2\phantom{ZZ}}{\delta I^{a}_\mu\delta K^{b}_\nu}
\right).
\nonumber
\end{align}
Operators acting on $\delta^{ab}$
are meant as differentiation with repect to $x_a$.
According to DeWitt's condensed notation \Eqref{eq:GonZ}
comprehends functional traces as usual.

Rewriting \Eqref{eq:flowZ} in terms of the Schwinger functional and performing the Legendre transformation to $\tGamma[\Phi,v,\cI]$, cf. \Eqref{eq:deftGamma}, we obtain the RG flow of the Legendre effective action,
\begin{widetext}
	\begin{align}
	\bs
	\partial_t\tGamma&=
    \frac{1}{2}\left(
	\partial_t R_{\mu\nu}\delta^{ab}\right)
	\left(
	\big(\tGamma^{(2)-1}\big)_{A^{a}_\mu A^{b}_\nu}
	+\tGamma^{(2)}_{M^a_\mu \Phi^j}
	\big(\tGamma^{(2)-1}\big)_{\Phi^j A^b_\nu}
	+\tGamma^{(2)}_{K^{b}_\nu I^{a}_\mu}
	+\frac{\delta\tGamma}{\delta M^{a}_\mu }A^{b\nu}
	-\bigg(\!\frac{\delta\ }{\delta K^{b}_\nu }\tGamma\bigg)
	\bigg(\!\mathop{\tGamma\frac{\delta\ }{\delta I^{a}_\mu }}^{\leftarrow}\!\bigg)
	\right)
	\\
	&\ \ \ +	\left( \partial_t r_\gh\partial_\mu\delta^{ab}\right) 
	\left(
	\tGamma^{(2)}_{K^a_\mu \Phi^j}
	\big(\tGamma^{(2)-1}\big)_{\Phi^j(-\bC^b)}
	-\bC^b \bigg(\!\frac{\delta\ }{\delta K^{a}_\mu }\tGamma\bigg)
	\right)
    +\partial_t \Delta S_\gf\, ,
	\label{eq:GontGamma}
	\es
	\end{align}
\end{widetext}
being an exact equation of (up to) one-loop structure. The
corresponding equation for the effective average action follows
straightforwardly upon substitution of \Eqref{eq:defGamma}, yielding a
Wetterich-type equation adapted to the present construction.

Notice that the last term of \Eqref{eq:GontGamma}
\begin{equation}
\partial_t \Delta S_\gf=
\left(\partial_t r_\gh\partial_\mu\delta^{ab}\right) 
v^b A^{a\mu}
+\frac{1}{2}
\left(\partial_t R_{\mu\nu}\delta^{ab}\right) A^{a\mu} A^{b\nu},
\end{equation}
cancels an identical contribution
on the left-hand side, whose presence is required
by the master equation, as argued in the next sections.

A deeper insight into the structure of the remaining terms can
be gained by considering the following class of truncations
\begin{align}
\bs
\tGamma_k [\Phi,v,\cI]= \tGamma[\Phi,v]+S_\so^{\mathrm{brst}}[\Phi,v,\cI]\,.
\label{eq:splittGamma}
\es
\end{align}
Here, we have assumed that the $\cI$ dependence 
 takes a simple linear form as in the bare action,
see \Eqref{eq:Ssobrst}.
This directly translates into a similar truncation
for the effective average action
\begin{align}
\bs
\Gamma_k [\Phi,v,\cI]= \Gamma[\Phi,v]+S_\so^{\mathrm{brst}}[\Phi,v,\cI]\,.\label{eq:splitGamma}
\es
\end{align}
For this family of truncations one finds
\begin{align}
\tGamma^{(2)}_{K^{a}_\nu I^{a}_\mu}&=0\\
\partial_t \Delta S_\gh&=
-\left( \partial_t r_\gh\partial_\mu\delta^{ab}\right)
\bC^b \bigg(\!\frac{\delta\ }{\delta K^{a}_\mu }\tGamma\bigg)
\\
+\frac{1}{2} & 
\left(\partial_t R_{\mu\nu}\delta^{ab}\right)  \left(
\frac{\delta\tGamma}{\delta M^{a}_\mu }A^{b\nu}
-\bigg(\!\frac{\delta\ }{\delta K^{b}_\nu }\tGamma\bigg)
\bigg(\!\mathop{\tGamma\frac{\delta\ }{\delta I^{a}_\mu }}^{\leftarrow}\!\bigg)
\right) .\nonumber
\end{align}

\begin{figure}[t!]
	\includegraphics[width=\columnwidth]{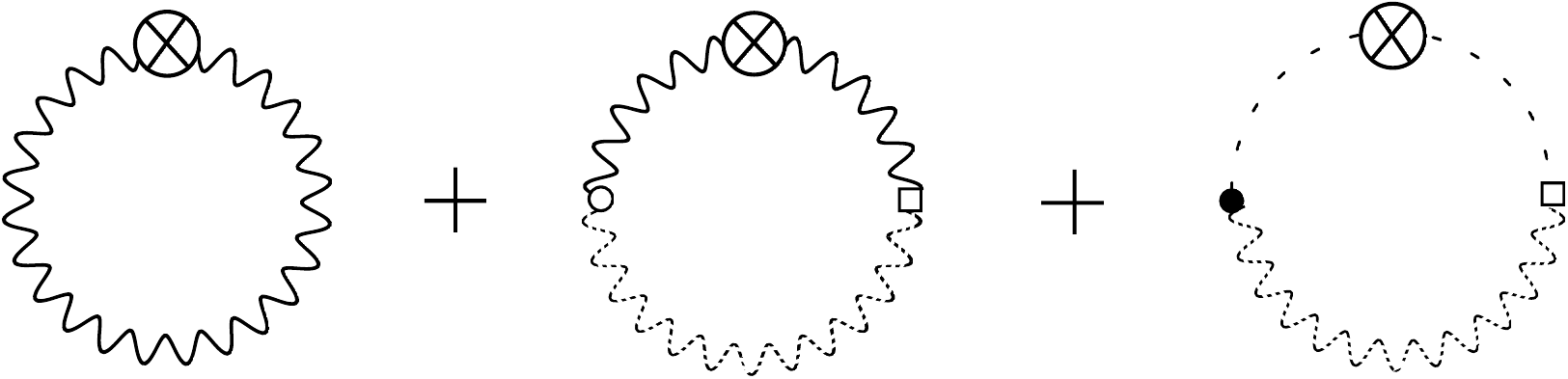}
	%
	\caption{Diagrammatic representation of the approximate flow equation in
		\Eqref{eq:GonGamma_approx}.}
	\label{fig:floweq}
\end{figure}

Thus, according to Eqs.~\eqref{eq:defGamma} and \eqref{eq:defDeltaS},
the flow of the source-independent part of the effective average action,
within the present truncation, obeys
\begin{align}
\partial_t\Gamma[\Phi,v]&=
\frac{1}{2}
\left( \partial_t R_{\mu\nu}\delta^{ab}\right)
\big(\tGamma^{(2)-1}\big)_{A^{a}_\mu A^{b}_\nu}
\nonumber\\
&+\frac{1}{2}
\left( \partial_t R_{\mu\nu}\delta^{ab}\right)
\tGamma^{(2)}_{M^a_\mu \Phi^j}
\big(\tGamma^{(2)-1}\big)_{\Phi^j A^b_\nu}
\label{eq:GonGamma_approx}
\\
&+
\left( \partial_t r_\gh\partial_\mu\delta^{ab}\right)
\tGamma^{(2)}_{K^a_\mu \Phi^j}
\big(\tGamma^{(2)-1}\big)_{\Phi^j(-\bC^b)}.
\nonumber
\end{align}
These three one-loop terms can be diagrammatically represented as in
\Figref{fig:floweq}.
Here, the crossed insertions represent the regulators:
either $\partial_t R_{\mu\nu}$ or $\partial_t r_\gh$
depending on whether they enter gluon or ghost lines respectively.
The full circle here represent the $v$-independent vertex
$\tGamma^{(2)}_{K^a_\mu \Phi^j}$. Later on 
we will use full circles to denote also
vertices arising from differentiation
of $\tGamma^{(2)}_{K^a_\mu \Phi^j}$
with respect to other copies of $\Phi$.
The empty circle represents the $v$-dependent
vertex $\tGamma^{(2)}_{M^a_\mu \Phi^j}$.
Again, in the following we will use empty 
circles also for vertices with more than two legs,
arising from further differentiation with respect to
$\Phi$.
Wavy lines correspond to gluons, dashed lines to ghosts,
and wavy-dashed lines to any of these.
Let us stress that \Eqref{eq:GonGamma_approx}
 is only an approximation of the
exact flow in \Eqref{eq:GontGamma}, as a nonlinear dependence of $\tGamma$ on
$\cI$ is generically expected.

%
%

\section{RG flow and the master equation}
\label{sec:flowmasterequation}

 Using the standard reasoning outlined in \Secref{sec:noise}, we derive the master equation from a BRST variation of the fields under the integral in \Eqref{eq:frgpartition}, assuming BRST invariance of the measure. We write the master equation as
\begin{equation}
\Sigma=0\,,
\label{eq:sigmaiszero}
\end{equation}
where $\Sigma$ arises from the BRST variations of the source terms,
\begin{align}
\bs
\Sigma[W]=&\,-J_\mu^a\frac{\delta W}{\delta K^a_\mu}
+\bar{\eta}^a\frac{\delta W}{\delta L^a}
+v^a\eta^a\\
&\,- M_\mu^a \frac{\delta W}{\delta K_\mu^a}
+\frac{\delta W}{\delta J_\mu^a}I^a_\mu
-\frac{\delta W}{\delta M_\mu^a}I_\mu^a\, .
\label{eq:masterW}
\es
\end{align}
Here, the functional dependence on $W$ is used only as an abbreviation for the dependence on all arguments of $W$. The last line arises from the BRST variations of the additional source terms. Now, we can straightforwardly perform the transformation to the Legendre 
effective action by using Eqs.~\eqref{eq:QoM} and \eqref{eq:WtoGamma} (with $\Gamma \to \tGamma$) and the new source relations
\begin{equation}
\frac{\delta \tGamma}{\delta M_\mu^a}= -\frac{\delta W}{\delta M_\mu^a}, \quad
\frac{\delta \tGamma}{\delta I_\mu^a}= -\frac{\delta W}{\delta I_\mu^a},
\label{eq:WtoGamma2}
\end{equation}
and arrive at
\begin{align}
\bs
\Sigma[\tGamma]=&\,\frac{\delta \tGamma}{\delta A_\mu^a}\frac{\delta \tGamma}{\delta K^{a\mu}}+\frac{\delta \tGamma}{\delta c^a}\frac{\delta \tGamma}{\delta L^a}
+v^a\frac{\delta \tGamma}{\delta\bar{c}^a}\\
&+ M_\mu^a \frac{\delta \tGamma}{\delta K_\mu^a}
+{A_\mu^a}I^{a\mu}+\frac{\delta \tGamma}{\delta M_\mu^a}I_\mu^a.
\label{eq:mastergamma}
\es
\end{align}
The first line of \Eqref{eq:mastergamma} together with \Eqref{eq:sigmaiszero} is identical to the Zinn-Justin master equation \eqref{eq:mastereqGamma}, representing the standard BRST symmetry constraint for the (Legendre) effective action as familiar from a linear gauge fixing. The second line represents new terms arising from the additional sources to take care of the regulator dependent vertices. Eqs.~\eqref{eq:mastergamma}, \eqref{eq:sigmaiszero} represent a main result of this work, as they encode the BRST symmetry of the scale-dependent regularized effective action on the same algebraic cohomology level as the conventional Zinn-Justin master equation. In contrast to the master equation derived in Ref.~\cite{Ellwanger:1994iz,Reuter:1994zn,DAttanasio:1996tzp}, \Eqref{eq:mastergamma} contains no loop terms and thus may lead to substantial technical simplifications also for nonperturbative approximation schemes. In fact, master equations remaining on the same algebraic level have previously been found in full generality for scale-dependent 2PI effective actions \cite{Pawlowski:2005xe}; for a special version of this construction, see \cite{Lavrov:2012xz}. Here, we obtain this property already on the level of a scale-dependent 1PI effective action.

We now need to proof the compatibility of the master equation with the
corresponding RG flow. For this let us assume that the kernels
$R_{\Lo}$, $R_{\T}$ and $r_\gh$ completely regularize the theory and
introduce the single floating momentum scale $k$. The desired
compatibility can most conveniently be shown on the level of the flow
of the partition function $Z$ of \Eqref{eq:frgpartition}.  The flow
equation for $Z$ reads, cf. \Eqref{eq:flowZ},
\begin{equation}
\partial_t Z=\G_t Z\, 
\label{eq:flowZ2}
\end{equation}
with the generator $\G_t$ given in \Eqref{eq:GonZ}.
In order to solve the flow equation \Eqref{eq:flowZ2}
independently of the master equation, it is
of key importance to show that the two equations are
compatible, in the sense that the symmetry condition
of \Eqref{eq:sigmaiszero}
is preserved by the RG flow.
Indeed this is the case since
\be
\partial_t \Sigma[Z]=\G_t \Sigma[Z]\,.
\label{eq:SigmaFlow}
\ee
In other words, one can 
define the BRST generator appropriate for $Z$ or $W$ as
\begin{align}
\bs
\mathcal{D}=&\,-J_\mu^a\frac{\delta \phantom{W}}{\delta K^a_\mu}
+\bar{\eta}^a\frac{\delta \phantom{W}}{\delta L^a}
+v^a\eta^a\\
&\,- M^\mu_a \frac{\delta \phantom{W}}{\delta K_\mu^a}
+ I^a_\mu\frac{\delta \phantom{W}}{\delta J_\mu^a}
-I^\mu_a\frac{\delta \phantom{W}}{\delta M^\mu_a}\, ,
\label{eq:DonW}
\es
\end{align}
such that $\Sigma[Z]=\mathcal{D}Z$.
Since the transformation in \Eqref{eq:BRSTagain} is nilpotent
one has $\mathcal{D}^2=0$.
Then one can show that $\mathcal{D}$ commutes
with $\G_t$
\be
\left[\mathcal{D},\G_t\right]=0\,.
\label{eq:commutator}
\ee
Details on the proof of this relation 
are given in \Appref{app:compatibility}.
Needless to say, compatibility
for $Z$ is equivalent to compatibility
for $W$ or $\tGamma$.
This compatibility as expressed through \Eqref{eq:SigmaFlow} directly
implies that a solution, say for the Legendre effective action
$\tGamma_k$, to the RG flow \eqref{eq:GontGamma} satisfies the master
equation $\Sigma=0$ on all scales $k$, provided its initial condition 
satisfies the master equation, say at an initial scale $k=\Lambda$.

\section{RG initial conditions - the reconstruction problem}
\label{sec:reconstruction}

The solution to the flow equation \eqref{eq:GontGamma} is equivalent
to the construction of the effective action via the functional
integral, provided the initial conditions are appropriately related to
the bare action entering the functional integral. The identification
of the initial conditions and their relation to the bare action is
known as the \textit{reconstruction problem}~\cite{Manrique:2008zw,Manrique:2009tj,Vacca:2011fx,Morris:2015oca}.

Let us thus address the behavior of the functional integral
in~\Eqref{eq:frgpartition} when the RG scale $k$ approaches the
asymptotic IR and UV limits.  As discussed in
\Secref{sec:regularization}, in the IR we have $k\to0$, and both
kernels $R^{\mu\nu}$ and $r_\gh$ then vanish.  Thus, the
scale-dependent effective action $\tGamma$ as defined through the
functional integral reduces to the full gauge-fixed effective action
in this limit.  Notice that this process does not reproduce the
standard Gaussian implementation of Lorenz gauge, but the
$\bar{m},\bar{m}_\gh\to0$ limit of the nonlinear gauge described in
\Secref{sec:nonlinearF}.

On the other hand, we would like the
UV limit to correspond to the case in which the
fields become infinitely massive, 
i.e.~$\bar{m}\to+\infty$ and $\bar{m}_\gh\to+\infty$.
In this case, the UV limit of the action is characterized by a decoupling of all modes.
As stated in \Secref{sec:regularization}
this can either occur when $k\to\Lambda$,
with $\Lambda$ being an independent UV cutoff,
or when $k\to+\infty$.
We choose the second option, for definiteness,
and we further assume the following behavior:
\begin{align}
\bs
R^{\mu\nu}(\partial)
&\widesim[2]{k\to+\infty}k^2\delta^{\mu\nu}\,,\\
r_\gh (-\partial^2)
&\widesim[2]{k\to+\infty}\frac{k^2}{-\partial^2}\,.
\es
\label{eq:asymptoticreg}
\end{align}
This implies that both the gluons and the ghosts
acquire the same mass in the UV limit.
While this is not the only possible scenario,
the following arguments
can be straightforwardly 
adjusted to different UV asymptotics.

%
%

To compute the $k\to+\infty$ limit of the
effective action from the functional integral,
we need
distinct notations for the fluctuating fields
inside the functional integral, 
and for the average ones.
As the latter have been collectively
grouped in the variable $\Phi$
in \Secref{sec:EAA},
we here introduce the notation
$\varphi$ to indicate the fluctuating
fields.
Then, we can write the functional integral
of \Eqref{eq:frgpartition}
in terms of the effective action as
\begin{equation}
e^{-\tGamma[\Phi,v,\mathcal{I}]}=\int\!\DD \varphi^i\,
e^{-S[\varphi,v]-S_\so^\brs[\varphi,v,\mathcal{I}]+\tGamma \frac{\overleftarrow{\delta}\ }{\delta\Phi^i}\left(\varphi-\Phi\right)^i }\, .
\label{eq:frgpartitiontGamma}
\end{equation}
Here, $S[\varphi,v]$ is the action defined in \Eqref{eq:bareS}
while $S_\so^\brs[\varphi,v,\mathcal{I}]$
has been introduced in \Eqref{eq:Ssobrst}.

As the regulators diverge, the bare action also
diverges, and the functional integral is dominated
by stationarity configurations. 
We can thus account for the $k\to+\infty$
limit by a
simple saddle-point approximation.
Thus, we first look for the maxima of the Euclidean
action inside the functional integral.
Clearly the crucial role is played by the diverging
parts of this action. As such,
we need specific preliminary assumptions 
to identify these parts.
In particular, we are interested in
the possibility that the effective action
$\Gamma$ and its derivatives,
as well as the average fields $\Phi^i$
and the sources $v^a$ and $\mathcal{I}_j$ stay finite
in the UV limit. 
That $\Gamma$ should stay finite instead of $\tGamma$,
is suggested by the BRST symmetry itself, i.e.~by
the master equation,
which forces  $\tGamma$ 
to comprehend the gauge-fixing and ghost actions.
 Thus, the diverging parts of the latter
 actions, as given in Eqs.~\eqref{eq:DeltaSgf}
and \eqref{eq:DeltaSgh},
must appear in both the bare action $S$
and in the effective action $\tGamma$.
We therefore inspect the functional integral
representation for $\Gamma$ descending from 
\Eqref{eq:frgpartitiontGamma},
namely
\begin{align}
\bs
&e^{-\Gamma[\Phi,v,\mathcal{I}]}=\int\!\DD \varphi^i\,
\exp\bigg\{-S[\varphi,v]-S_\so^\brs[\varphi,v,\mathcal{I}]\\
&+\Delta S[\Phi,v]
+\Delta S\!\!\mathop{\ \frac{{\delta}\ }{\delta\Phi^i}}^\leftarrow
\left(\varphi-\Phi\right)^i 
+\Gamma\!\! \mathop{\ \frac{{\delta}\ }{\delta\Phi^i}}^\leftarrow \left(\varphi-\Phi\right)^i 
\bigg\}\, .
\es
\label{eq:frgpartitionGamma}
\end{align}
where $\Delta S$ has been defined in \Eqref{eq:defDeltaS}.
In the last functional integral, the diverging 
contributions to the Euclidean action read
\begin{align}
S_\mathrm{div}=\Delta S[\varphi,v]-\Delta S[\Phi,v]
-\Delta S[\Phi,v] \!\!\mathop{\ \frac{{\delta}\ }{\delta\Phi^i}}^\leftarrow\left(\varphi-\Phi\right)^i \,.
\end{align}
The saddle point configuration for $\varphi$ can then be 
otained by solving the condition
\begin{align}
 S_\mathrm{div}\!\! \mathop{\ \frac{{\delta}\ }{\delta\varphi^i}}^\leftarrow
 = \Delta S[\varphi,v]\!\!\mathop{\ \frac{{\delta}\ }{\delta\varphi^i}}^\leftarrow
-\ \Delta S[\Phi,v] \!\!\mathop{\ \frac{{\delta}\ }{\delta\Phi^i}}^\leftarrow
=0\,.
\label{eq:stationarity}
\end{align}
This admits the solution $\varphi^i=\Phi^i$.
In order for the $k\to+\infty$ limit to
produce a functional delta $\delta[\varphi^i-\Phi^i]$,
this configuration must be the unique solution
and correspond to a maximum.
Thus we have to check that the operator
\begin{equation}
\frac{\delta\ }{\delta \varphi^{\dagger\,j}} 
S_\mathrm{div}\!\! \mathop{\ \frac{{\delta}\ }{\delta\varphi^i}}^\leftarrow
= 
\frac{\delta\ }{\delta \varphi^{\dagger\,j}}
\Delta S[\varphi,v]\!\!\mathop{\ \frac{{\delta}\ }{\delta\varphi^i}}^\leftarrow
=\Delta S^{(2)}_{ji}[\varphi,v]
\label{eq:positive}
\end{equation}
be positive definite.

To inspect the explicit component form of 
Eqs.~\eqref{eq:stationarity} and \eqref{eq:positive},
we take advantage of the assumptions
in \Eqref{eq:asymptoticreg}.
Then we can write
\begin{align}
\frac{1}{k^2}\frac{\delta \Delta S }{\delta A^{a\mu}}&=
A^a_\mu+\frac{\partial_\mu}{\partial^2} v^a
- \bar{g}f^{bca}
\left(\frac{\partial_\mu}{\partial^2}\bC^b\right)
 c^c
 -\frac{v^b\bC^b}{|v|^2}
\partial_\mu c^a
\,,
\label{eq:stationarityA}\\
\frac{1}{k^2}\frac{\delta\ }{\delta \bC^a}\Delta S&=
c^a+\bar{g}f^{abc}
\frac{\partial_\mu}{\partial^2}
\left(A^c_\mu c^b\right)
 -\frac{v^a}{|v|^2}
A^{b\mu}\partial_\mu c^b
\,.
\label{eq:stationarityc}
\end{align}
A similar equation can be obtained for the derivative
with respect to $c^a$, which is not needed for our 
discussion.
We observe that the second term on
the right-hand side of \Eqref{eq:stationarityA}
cancels in the difference on the right-hand side of
\Eqref{eq:stationarity},
such that possible solutions to
the stationarity condition different from $\varphi^i=\Phi^i$
must correspond to
gluon configurations which involve the expectation values
of the ghosts.
Also, as the mixed derivatives in the fluctuation operator
of \Eqref{eq:positive} are nonvanishing,
assessing the positiveness of the latter is 
a nontrivial task.
Furthermore, while the gluon-gluon diagonal
block of this matrix is trivially positive,
the ghost-antighost diagonal block is
just (the diverging part of) the Faddeev-Popov operator.
Therefore, possible violations of positivity of the matrix in \Eqref{eq:positive}
are closely related to the Gribov ambiguity.
We comment more extensively on the latter issue in \Secref{sec:conclusions}.

We nevertheless argue that the requested positivity
must hold at least in the perturbative 
regime of infinitesimal field amplitudes,
i.e.,~expectation values.
In fact, in this case the nonlinear terms
in the Eqs.~\eqref{eq:stationarityA}
and \eqref{eq:stationarityc} can be neglected,
such that the solution $\varphi^i=\Phi^i$
becomes unique and \Eqref{eq:positive} reduces
to a positive mass matrix.
Thus, for infinitesimal field amplitudes we
do obtain a rising delta functional 
$\delta[\varphi^i-\Phi^i]$ in the $k\to+\infty$
limit of \Eqref{eq:frgpartitionGamma}, provided we introduce a $k$-dependent normalization of the functional measure~\cite{Wetterich:1989xg,Kopietz:2010zz,Vacca:2011fx} to guarantee
\begin{equation}
  \lim_{k\to\infty} \int d\mu[\varphi,v]\, \exp \big( -S_{\text{div}} \big) = \delta[\varphi^i-\Phi^i]\,.
  \label{eq:measurerequirement}
\end{equation}
In the present case the measure required for \Eqref{eq:measurerequirement}
reads
\begin{equation}
\mu[\varphi,v]=\left(
\mathrm{Det}\, \Delta S^{(2)}[\varphi,v]
\right)^{\frac{1}{2}}\,.
\end{equation}
Notice, however, that this measure
is field-dependent beyond the limiting case of
infinitesimal field amplitudes.
As such, this choice of measure
in the functional integral would 
bring nontrivial contributions to the flow
equation, which we have not included
in \Eqref{eq:GontGamma}.
To preserve the simple form of
\Eqref{eq:GontGamma},
we should instead adopt a field-independent
normalization of the functional measure,
for instance $\mu[0,v]$.
%
%
In this case the $k\to+\infty$ limit is
finite and nontrivial and reads
\begin{equation}
\lim_{k\to \infty} e^{-\Gamma[\Phi,v,I]}=
e^{-S[\Phi,v,I]}
\lim_{k\to \infty}\frac{\mu[0,v]}{\mu[\Phi,v]}.
\end{equation}
The correction 
to the bare action encoded in the ratio of measures on the right-hand side 
of the last equation is expected 
whenever the regularization 
of the functional integral is 
more then quadratic in the fluctuating
fields,
and a field-independent functional measure is adopted.
Indeed the latter has been observed also
in a similar symmetry-preserving 
functional-renormalization-group formulation
of nonlinear sigma models~\cite{Vacca:inprep}.
For a recent discussion of measure or normalization induced terms
for background-field flows, see \cite{Lippoldt:2018wvi}.

Thus, the task of constructing initial conditions
$\Gamma_{k=\Lambda}$
appropriate for the chosen bare action $S$
requires evaluation of the counterterm action
\begin{align}
S_\mathrm{c.t.}=\frac{1}{2}\Tr \log
\Delta S^{(2)}[\Phi,v]
-\frac{1}{2}\Tr \log
\Delta S^{(2)}[0,v]
\label{eq:counterterm}
\end{align}
at the initialization scale $k=\Lambda$,
such that
\begin{align}
\Gamma_{k=\Lambda}=S+S_\mathrm{c.t.}\,.
\end{align}
The evaluation of the trace in \Eqref{eq:counterterm}
requires approximations and truncations similar to 
those employed for solving the flow equation.
The main open question regarding the latter task
is whether this functional trace is regularized.
In fact, while the flow equation involves the 
derivative of the regulators with respect to $t=\log k$,
\Eqref{eq:counterterm} features only $R$ and $r_\gh$ 
at the scale $k=\Lambda$.
The requirement that $S_\mathrm{c.t.}$ be finite
within the chosen truncation, might bring 
novel constraints on the allowed regulators.
A detailed analysis of these issues is deferred to
future works.

\section{The leading-order gluon wave-function renormalization}
\label{sec:oneloop}

We illustrate the new flow equation \eqref{eq:GontGamma}
with the simple application of
computing the gauge-boson 
 wave-function renormalization.
To this end, we use an ansatz
for the 1PI effective action
within the family of \Eqref{eq:splittGamma},
i.e.,~a functional linear in the sources.
It is sufficient to further specialize also the rest of
the effective action to be of the bare form,
but with scale-dependent parameters,
\begin{align}
\bs
\tGamma_k [\Phi,v]=& Z_\T S_\YM[A]
+\tGamma_\gf[A]+\tGamma_\gh[A,c,\bC,v]\,.
\label{eq:baretGamma}
\es
\end{align}
The gauge-fixing and ghost actions now include 
ghost and gluon wave-function renormalizations
\begin{align}
&\tGamma_{\gf}= \frac{1}{2}{A}_\mu^a \X^{\mu\nu}\! A^{a}_\nu + v^a Z_\gh(1+r_\gh(-\partial^2)) \partial^\mu\! A^{a}_\mu\,,\label{eq:tGammagf}\\
\bs
&\tGamma_{\gh}= -Z_\gh\bar{c}^a (1+r_\gh(-\partial^2)) \left(\partial^\mu D_\mu c\right)^a 
\label{eq:tGammagh}\\
&-\frac{v^a}{2|v|^2}\bar{c}^a \left(
\left(\X^{\mu\nu}{A}^b_\nu\right) \left({D}_{\mu}c\right)^b
+{A}^b_\mu\left(\X^{\mu\nu}{D}_{\nu}c\right)^b 
\right).
\es
\end{align}
Furthermore, the quadratic kernel $\X$ must also
depend on the  wave-function renormalizations of
the longitudinal and transverse gluons.
This can be accounted for by 
choosing this kernel as in \Eqref{eq:frgX}
and \Eqref{eq:Rmunu}
and by rescaling 
$\xi\to\xi/Z_\Lo$ and $R_\T\to Z_\T R_\T$, 
$R_\Lo\to Z_\Lo R_\Lo$.
A similar rescaling has to be performed also in 
$\Delta S$ of \Eqref{eq:defDeltaS}.
Therefore we can write
the corresponding effective average action as
\begin{align}
\bs
\Gamma_k [\Phi,v]=& Z_\T S_\YM[A]
+\Gamma_\gf[A]+\Gamma_\gh[A,c,\bC,v]\,.
\label{eq:bareGamma}
\es
\end{align}
where now
\begin{align}
\Gamma_{\gf}& = \frac{Z_\Lo}{2\xi}\left(\partial^\mu{A}_\mu^a\right)^2  
+ Z_\gh v^a \partial^\mu\! A^{a}_\mu\,,\label{eq:Gammagf}\\
\Gamma_{\gh}& = -Z_\gh\bar{c}^a  \left(\partial^\mu D_\mu c\right)^a 
\label{eq:Gammagh}\\
&+\frac{Z_\Lo v^a}{2\xi|v|^2}\bar{c}^a \left(
\left(\partial^\mu\partial^\nu {A}^b_\nu\right)\left({D}_{\mu}c\right)^b
+{A}^b_\mu\left(\partial^\mu\partial^\nu {D}_{\nu}c\right)^b 
\right).\nonumber
\end{align}
As the truncation of \Eqref{eq:baretGamma}
has the same functional form as the bare action,
it trivially solves the master equation.
We can thus extract the running of the parameters
$Z_\T$, $Z_\Lo$, $Z_\gh$ and $\bar{g}$
from several different but equivalent operators.

We adopt Feynmann gauge for technical convenience, 
namely we evaluate the
right-hand side of the flow equation 
by choosing 
\begin{align}
\bs
 \xi&=1\,\\
 Z_\Lo&=Z_\T=Z \,,\\
 R_\Lo(p^2)&=R_\T(p^2)=R(p^2)\,.
 \label{eq:Feynmangauge}
 \es
\end{align}
This results in equal propagators for the longitudinal and for
the transverse gluons.
Furthermore, in this work we limit  ourselves to the
leading term in a perturbative expansion about $\bar{g}^2=0$.

\begin{figure}[t!]
	\includegraphics[width=\columnwidth]{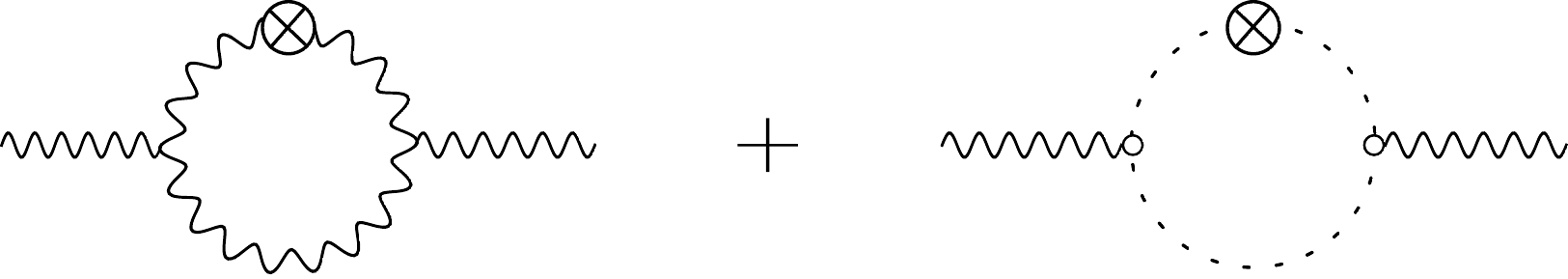}
	%
	\caption{Diagrams contributing to the gluon wave-function renormalization.}
	\label{fig:nonvanishing}
\end{figure}

The diagrams contributing to the gluon-wave function
renormalization are displayed in \Figref{fig:nonvanishing}.
The gluon loop is universal, and 
evaluates to a standard Feynman-gauge
one-loop result~\cite{Peskin:1995ev}.
There is no usual ghost loop
contributing through standard $v$-independent
vertices.
This fact
can be understood by observing that,
in absence of the background field $v$,
a simple rescaling of the ghost fields
could be used to remove the regulator $r_\gh$
from the ghost sector.
This suggests that all ghost-loop contributions
on the right-hand side of \Eqref{eq:GontGamma}
involve $v$-dependent vertices.
More details on this cancellation are 
provided in \Appref{app:computations}.

The only nonvanishing ghost-loop contribution 
comes from the second diagram 
in \Figref{fig:nonvanishing}.
Here, the empty circles represent the
$v$-dependent two-ghosts-one-gluon vertex
proportional to $\X$, which can be computed
from \Eqref{eq:tGammagh}.
As such, the result of this diagram cannot 
be regulator-independent, since $R$ 
defines the gauge fixing and brings new momentum-dependent vertices.
Thus, we must specialize our discussion to
a particular regulator choice.
For reasons of mathematical convenience,
we adopt the piecewise linear 
regulator~\cite{Litim:2001up}
for both the gluon and the ghost propagators.
In formulas, we choose 
\begin{equation}
R(q^2)=q^2 r(q^2)\,,\quad\quad r_\gh(q^2)=r(q^2)\,,
\label{eq:equalregulators}
\end{equation}
and
\begin{equation}
r(q^2)=\left(\frac{k^2}{q^2}-1\right)\theta(k^2-q^2)\,.
\label{eq:rlinear}
\end{equation}
Let us define dimensionless renormalized fields 
and couplings in $d=4$ as 
\begin{align}
g^2&=\bar{g}^2/Z\,,
\label{eq:renormg}\\
\tilde{v}^a&=\frac{Z_\gh\, \bar{g}}{Z k^2}v^a\,.
\label{eq:renormv}
\end{align}
Let us introduce projectors on the adjoint color subspaces
which are parallel or perpendicular to $v^a$
\begin{align}
\Pi_{\parallel}^{ab}=\frac{\tilde{v}^a \tilde{v}^b}{|\tilde{v}|^2}\,,\qquad \Pi_{\perp}^{ab}=\delta^{ab}-\frac{\tilde{v}^a \tilde{v}^b}{|\tilde{v}|^2}\,.
\label{eq:colorprojectors}
\end{align}
We also need to differentiate the wave function renormalizations
for colors parallel or perpendicular to $v$, by adding
corresponding superscripts.
The final result for a constant background field
$v^a(x)$ then reads
\begin{align}
\partial_t  Z_\T^{\perp}  &= \frac{Z}{16\pi^2}
 g^2 C_2(G) \frac{19}{6}\,,
\label{eq:finalZTperp}
\\
\partial_t  Z_\Lo^{\perp} &= - \frac{Z}{16\pi^2}  g^2 C_2(G) \frac{1}{2}
\,,
\label{eq:finalZLperp}
\\
 \partial_t  Z_\T^\parallel  &= \frac{Z}{16\pi^2} g^2 \left[
 C_2(G) \frac{19}{6}
-\frac{1}{4|\tilde{v}|^2}
\left(\frac{1}{6}+\frac{1}{4}\right)\right], 
\label{eq:finalZTpar}
\\
\partial_t  Z_\Lo^\parallel &= \frac{Z}{16\pi^2} g^2 \left[
- C_2(G) \frac{1}{2}
-\frac{1}{4|\tilde{v}|^2}
\left(\frac{1}{2}-\frac{1}{4}\right)\right]. 
\label{eq:finalZLpar}
\end{align}
Here, $C_2(G)$ is the Casimir in the adjoint representation.
The presence of a background field $v$,
which explicitly breaks the global symmetry,
entails that colors parallel or perpendicular
to $v$ renormalize differently.
While perpendicular colors
 receive contributions from gluon loops only,
the parallel color is affected by ghosts
as well.
The $v$-dependent contribution has been written 
in a form that allows for comparison with the standard
ghost-loop contribution in Feynman gauge.
In fact, the present ghost loop differs from the latter
in two ways: first, the factor $\delta^{ab} g^2 C_2(G)$
is replaced by $-\tilde{v}^a \tilde{v}^b/(4|\tilde{v}|^4)$;
second, the momentum dependence of the vertices is augmented 
by the presence of the regulators.
Ignoring the regulator contribution to the vertices would 
result in the first number within parentheses ($1/6$ and
$1/2$ for the transverse and longitudinal modes respectively),
which is the universal result in the usual Feynman gauge.
Thus, the second number ($1/4$ and
$-1/4$ for the transverse and longitudinal modes respectively)
is the contribution due to the presence of $R$ in the
vertices.

\section{Conclusions}
\label{sec:conclusions}

This work addresses the continuum formulation of quantum Yang-Mills
theory in $d=4$ Euclidean spacetime dimensions,
within a gauge-fixed functional approach.
It especially focuses on the issue of providing a regularization
of the corresponding quantum field theory in a mass-dependent RG scheme,
while preserving the underlying BRST symmetry.
This goal is achieved by means of a careful implementation of the 
gauge fixing, in a two-step process.
First, we depart from the most popular choice of performing a Gaussian
average over the noise field that is used to implement the action for
the Nakanishi-Lautrup field as part of the gauge-fixing sector. Instead, we introduce a background 
Nakanishi-Lautrup field $v^a$ by means of an imaginary exponential
distribution (Fourier weight) for the noise.
This results in a gauge-fixing action which is linear in the 
gauge-fixing functional $\F^a[A]$, cf.~\Secref{sec:noise}.

Second, we specialize to the case in which $\F^a[A]$ is
nonlinear in $A$, and in particular it comprehends both
a linear and a quadratic part.
As discussed in \Secref{sec:nonlinearF}, 
a nonvanishing  quadratic part of $\F^a$ can accommodate a mass parameter
$\bar{m}$ for the gluons which does not require any nonlocality
and is by construction compatible with BRST symmetry.
On the other hand, 
the linear part of $\F^a$ is desirable as it provides a quadratic kinetic 
term for the ghost fields. For definiteness, we choose it such that
it reproduces the standard ghost action in Lorenz gauge.
This linear term in the gluon sector can then be interpreted as a source action
for $\partial^\mu A^a_\mu$, with the field $v^a$ as the corresponding source.
In \Secref{sec:nonlinearF}, we also notice that one can take advantage of the linear term in $\F^a$ 
to include a ghost mass parameter $\bar{m}_\gh$ in the ghost action.
Though the latter appears to provide an IR regularization of the bare ghost propagator,
it has the simultaneous effect of introducing nonlocal ghost-gluon vertices.
This is a direct consequence of the fact that BRST symmetry, similarly to gauge symmetry, mixes modes nonlocally in momentum space.
It is thus not clear whether such an $\bar{m}_\gh$ can be helpful in perturbative computations along the lines as suggested in various phenomenological or conceptual approaches to gauge theories.

In the present work, we take advantage of this particular gauge-fixing strategy to
construct a manifestly BRST-invariant functional renormalization group representation
of quantum Yang-Mills theory.
For this, we generalize the gauge-fixing to the case of momentum-dependent
mass parameters, cf.~\Secref{sec:FRGconstruction}, also 
%
%
introducing a floating RG scale $k$ in the bare action.
Considering the dependence of the generating functional on $k$ then leads to an exact RG flow equation
for the partition function and correspondingly for the 1PI effective action.
Though the regulator functions enter the ghost gluon vertices,
we observe that it is possible to put the flow equation into a one-loop
form, involving only second order derivatives of the 1PI effective action -- analogously to the Wetterich equation.
This is achieved by means of some sources for composite operators.
Two of them representing $a$-number sources $K^a_\mu$ and $L^a$ are already familiar from
the Zinn-Justin treatment of the master equation, as they couple to BRST variations
of the fields.
We introduce two additional $c$-number sources, $I^a_\mu$ and $M^a_\mu$ for the purpose of 
simplifying the flow equation.
The final result is presented in \Eqref{eq:GontGamma}.

The BRST symmetry of the bare action 
leads to a master equation \eqref{eq:mastergamma}, for the scale-dependent 
effective action,
which is a standard Zinn-Justin equation,
augmented with terms corresponding to the
new sources $I^a_\mu$ and $M^a_\mu$.
It can be solved algebraically
with the help of BRST cohomology.
The master equation is compatible with the flow equation,
as explained in \Secref{sec:flowmasterequation}
and proved in \Appref{app:compatibility}.
Therefore, if the effective action
fulfills the master equation at some scale,
it does so at all scales.
This is the case also for the standard functional-RG implementation, resulting in  modified
Ward-Takahashi and Slavnov-Taylor identities.
However, in the modified case, the
presence of the regulator in the symmetry 
identity makes it difficult in practice to
construct approximations which preserve this
compatibility.
This is not so in the present case.
Functional truncations satisfying the master
equation can be easily constructed by imposing 
manifest BRST symmetry, and then inserted 
into the flow equation. Compatibility
then entails that these truncations 
remain BRST symmetric along the flow.

The fact that the RG flow equation is exact,
does not necessarily mean that it 
usefully encodes the complete scale dependence
(momentum dependence) of correlation functions.
The latter point is crucially related 
to further important and mutually related requirements: 
1) that the
flow is fully regularized and no 
residual divergences remain;
2) that the regularization corresponds
to a physical coarse-graining,
with well-defined full-decoupling and 
full-propagation limits.
The second property is easier to assess in full generality than the first.
We have presented arguments in \Secref{sec:reconstruction} in support of the second requirement. 
In particular, we outline a concrete construction scheme to address the so-called reconstruction problem
of the bare action from the effective action,
i.e.,~the existence of a controlled full-decoupling
limit. 

Addressing the first requirement in a systematic
fashion is more involved. In the present work, we approach this question
by way of example, performing a perturbative computation of the
gluon wave-function renormalization. No residual divergences appear
in this case. 
On the contrary, a quite generic mechanism for the cancellation of 
possible IR divergences in the ghost sector is 
unveiled.
At first sight, ghost loops seem to  have the tendency to
show IR divergences, as the
regularization brought by $R_\gh$ is
multiplicative, affecting both
kinetic terms and vertices, therefore naively suggesting the existence of
unregularized diagrams. By contrast, we find that the sum of these dangerous diagrams
vanishes. Our results in
\Secref{sec:oneloop} suggest a general explanation
for this fact.

Our formalism requires the inclusion of an external Nakanishi-Lauptrup
field $v^a$. Taken at face value, this field -- if taken as given and
fixed -- is a source of explicit global color symmetry
breaking. However, since $v^a$ is introduced as part of the
gauge-fixing sector, we expect it to not contribute to any physical
observable. Moreover, color charge conservation is preserved if $v^a$
satisfies a (regulator-dependent) equation of motion. In practice, it
might be useful to treat $v^a$ as a quenched disorder field. For
instance, using a global Gaussian-type disorder with a suitably adjusted
(complex) width, the disorder average of the results for the universal parts of
the gluon wave-function renormalization corresponds to those of a
standard Feynman gauge fixing.

More extensive computations are needed to
further test the absence or presence
of residual divergences, and to explore
the properties of the RG flows generated
by this formalism. 
The full analysis of
the RG flow of the simplest 
perturbatively renormalizable truncation
of Eqs. \eqref{eq:bareGamma}, \eqref{eq:Gammagf}
and \eqref{eq:Gammagh}
is subject to future works.
For practical applications, generalizations may be useful that include
a background gauge field. 

Finally, it is worthwhile mentioning that our approach may allow for a
new perspective onto the Gribov problem, i.e., the problem of the
existence of multiple solutions of the gauge-fixing condition. Since
the scale-dependent regularization enters the construction via the
gauge fixing, also the relation between different Gribov copies if
they exist assumes a scale-dependent form. In the general case, also a
nonlinear gauge fixing such as ours is expected to permit the
existence of Gribov copies. This is visible, e.g., in
\Eqref{eq:SgfmasslikeTL} where terms of opposite signs occur in the
gauge-fixing condition, allowing for various forms of
cancellations. However in the strict Landau-gauge limit, we have
argued that the gauge-fixing condition cannot be satisfied for massive
modes. Replacing the mass parameter by a (strictly non-negative)
momentum-dependent regulator term, makes it clear that the regulator
takes influence on the existence of Gribov copies: for instance, a
would-be Gribov copy of a high-momentum massless transversal mode is
pushed away from the gauge orbit, if it entails 
 low-momentum
modes that cannot satisfy the gauge condition
due to the regulators. 
In other words, in the limiting case when regulator terms dominate 
	the gauge fixing functional,
	 any zero-mode of the Faddeev-Popov operator
	would be completely unregularized, a situation
	which is forbidden at any nonvanishing floating scale $k$ for
	a well-posed and smooth functional IR regularization.
This mechanism could
alleviate the Gribov problem in our construction. On the other hand,
it is clear that all Gribov copies, say of the Landau gauge, will reappear
in the limit $k\to0$, when the regulator is removed. A proper
discussion of the Gribov problem in our construction hence requires a
careful analysis of the various limits.

\section*{Acknowledgments}

We thank Riccardo Martini, Jan Pawlowski, Christof Wetterich and Omar
Zanusso for useful discussions. LZ acknowledges collaboration with
Gian Paolo Vacca on related projects. This work received funding
support by the DFG under Grants No. GRK1523/2 and No. Gi328/9-1.

\appendix

\section{Momentum-space conventions}
\label{app:Fourier}

The Fourier transform of field variables reads
\begin{equation}
\Phi^i(x)=\int_q \Phi^i(q)e^{\I qx}\,,
\end{equation}
where 
\begin{equation}
\int_q=\int\!\frac{\dd^d q}{(2\pi)^d}\,.
\end{equation}
Then \Eqref{eq:W1=Phi} can be written in momentum space as
\begin{equation}
\frac{\delta\ }{\delta \cJ_i^\dagger(-p)}W=\Phi^i(p)\,,\quad\quad
W \mathop{\frac{\delta\ }{\delta \cJ_i(p)}}^{\leftarrow}=\Phi^{\dagger i}(-p)\,,
\label{eq:W1=Phi_Fourier}
\end{equation}
and similarly for \Eqref{eq:Gamma1=J}.
The second-order derivatives of \Eqref{eq:W2} and \Eqref{eq:Gamma2}
are correspondingly defined as 
\begin{align}
W^{(2)}_{\cJ_i \cJ_j}
&=\frac{\delta\ }{\delta \cJ_i^\dagger(-p_1)}W\mathop{\frac{\delta\ }{\delta \cJ_j(p_2)}}^{\leftarrow}\,,
\label{eq:W2_Fourier}\\
\tGamma^{(2)}_{\Phi^i \Phi^j}
&=\frac{\delta\ }{\delta \Phi^{\dagger i}(-q_1)}\tGamma
\mathop{\frac{\delta\ }{\delta \Phi^j(q_2)}}^{\leftarrow}\,.
\label{eq:Gamma2_Fourier}
\end{align}
Thus, $\Phi^\dagger$ or $\cJ^\dagger$ are always evaluated at
reflected momenta.
This extends to all formulas, for instance to
Eqs.~\eqref{eq:W2II}, \eqref{eq:W2JI} and \eqref{eq:W2IJ}.
In particular, the momentum-space form of the flow equation,
\Eqref{eq:GontGamma}, reads
\begin{widetext}
	\begin{align}
	\partial_t\tGamma&=
	\frac{1}{2}\int_{q_1,q_2}
	\partial_t R_{\mu\nu}(q_2)\delta(q_1-q_2)
	\Bigg[
	\frac{\delta\tGamma}{\delta M^{a}_\mu(q_1)}A^{a\nu}(q_2)
	-\bigg(\!\frac{\delta\ }{\delta K^{a}_\nu(-q_2) }\tGamma\bigg)
	\bigg(\!\mathop{\tGamma\frac{\delta\ }{\delta I^{a}_\mu(q_1) }}^{\leftarrow}\bigg)
	+\tGamma^{(2)}_{K^{a}_\nu(-q_2) I^{a}_\mu(q_1)}
	\nonumber
	\\
	&
	\phantom{+\frac{1}{2}\int_{q_1,q_2}
		\partial_t R_{\mu\nu}(q_2)\delta(q_1-q_2)+}
	+\big(\tGamma^{(2)-1}\big)_{A^{a}_\mu(-q_2) A^{a}_\nu(q_1)}
	+\int_s\tGamma^{(2)}_{M^a_\mu(q_1) \Phi^j(s)}
	\big(\tGamma^{(2)-1}\big)_{\Phi^j(-s)A^a_\nu(-q_2)}\,
	\bigg]
	\label{eq:GontGamma_Fourier}
	\\
	&
	+\int_{q_1,q_2}
	\partial_t r_\gh(q_2^2)\, \I q_{2\mu}\delta(q_1-q_2)
	\Bigg[
	-\bC^a(-q_1) \bigg(\!\frac{\delta\ }{\delta K^{a}_\mu(-q_2) }\tGamma\bigg)
	+\int_s \tGamma^{(2)}_{K^a_\mu(-p_2) \Phi^j(s)}
	\big(\tGamma^{(2)-1}\big)_{\Phi^j(-s)(-\bC^a(q_1))}
	\Bigg]
	+\partial_t \Delta S_\gf,
	\nonumber
	\end{align}
\end{widetext}
and
\begin{align}
\bs
\partial_t \Delta S_\gf&=
\int_{q_1,q_2}
\partial_t r_\gh(q_2^2)\, \I q_{2\mu}\delta(q_1-q_2)
v^a(-q_1) A^{a\mu}(q_2)\\
&+\frac{1}{2}\int_{q_1,q_2}
\partial_t R_{\mu\nu}(q_2)\delta(q_1-q_2)
A^{a\mu}(-q_1) A^{a\nu}(q_2).
\es
\end{align}

\section{Compatibility proof}
\label{app:compatibility}


In this appendix, we provide more details
of the proof of \Eqref{eq:commutator}, forming the core of the compatibility proof.
This requires the cancellation of
a few nontrivial terms,
which appear when commuting $\DD$,
as defined in \Eqref{eq:DonW},
with the single functional derivatives
which additively contribute to $\G_t$,
see \Eqref{eq:GonZ}.

Let us start with the ghost loop,
i.e.,~with the derivatives contributing
to the functional trace which is 
regularized by $\partial_t r_\gh$.
When applied to $\DD$, they give
\begin{align}
v^b\frac{\delta\ }{\delta J^a_\mu}\DD&=
-v^b\frac{\delta\ }{\delta K^a_\mu}
+\DD\, v^b\frac{\delta\ }{\delta J^a_\mu}\,,\\
\frac{\delta\ }{\delta \eta^b}
\frac{\delta\ }{\delta K^a_\mu}\DD&=
-\frac{\delta\ }{\delta \eta^b}\DD
\frac{\delta\ }{\delta K^a_\mu}=
-v^b\frac{\delta\ }{\delta K^a_\mu}
+\DD \frac{\delta\ }{\delta \eta^b}
\frac{\delta\ }{\delta K^a_\mu}\,.
\end{align} 
As the operator $\G_t$ involves the 
difference of these two 
differential operators,
the nontrivial terms (i.e., the first terms) on the right-hand
sides cancel.

Next, we address the gluon loop
corresponding to  the second line of \Eqref{eq:GonZ}.
The latter involves three 
second-order functional derivatives.
Commuting each of them with $\DD$
gives
\begin{align}
\frac{\delta\ }{\delta J^a_\mu}\frac{\delta\ }{\delta J^b_\nu}\DD
&=\frac{\delta\ }{\delta J^a_\mu}\left(
-\frac{\delta\ }{\delta K^b_\nu}
+\DD \frac{\delta\ }{\delta J^b_\nu}\right)
\label{eq:deltaJJ_D}\\
&=-\frac{\delta\ }{\delta J^a_\mu}
  \frac{\delta\ }{\delta K^b_\nu}
 -\frac{\delta\ }{\delta K^a_\mu}
  \frac{\delta\ }{\delta J^b_\nu}
+\DD
\frac{\delta\ }{\delta J^a_\mu}\frac{\delta\ }{\delta J^b_\nu}
\,,\nonumber\\
\frac{\delta\ }{\delta M^a_\mu}\frac{\delta\ }{\delta J^b_\nu}\DD
&=\frac{\delta\ }{\delta M^a_\mu}\left(
-\frac{\delta\ }{\delta K^b_\nu}
+\DD \frac{\delta\ }{\delta J^b_\nu}\right)
\label{eq:deltaMJ_D}\\
&=-\frac{\delta\ }{\delta M^a_\mu}
\frac{\delta\ }{\delta K^b_\nu}
-\frac{\delta\ }{\delta K^a_\mu}
\frac{\delta\ }{\delta J^b_\nu}
+\DD
\frac{\delta\ }{\delta M^a_\mu}\frac{\delta\ }{\delta J^b_\nu}
\,,\nonumber\\
\frac{\delta\ }{\delta I^a_\mu}\frac{\delta\ }{\delta K^b_\nu}\DD
&=-\frac{\delta\ }{\delta I^a_\mu}
\DD \frac{\delta\ }{\delta K^b_\nu}
\label{eq:deltaIK_D}\\
&=-\frac{\delta\ }{\delta J^a_\mu}
\frac{\delta\ }{\delta K^b_\nu}
+\frac{\delta\ }{\delta M^a_\mu}
\frac{\delta\ }{\delta K^b_\nu}
+\DD
\frac{\delta\ }{\delta I^a_\mu}\frac{\delta\ }{\delta K^b_\nu}
\,.\nonumber
\end{align}
Taking \Eqref{eq:deltaJJ_D} minus \Eqref{eq:deltaMJ_D} 
minus \Eqref{eq:deltaIK_D}, 
which is the linear combination appearing in \Eqref{eq:GonZ},
results in the cancellation of the nontrivial commutator terms.
In summary, this verifies \Eqref{eq:commutator}.
Notice that this proof of compatibility does not require
the regulators to be diagonal in color indices,
as we have assumed throughout this work for reasons of convenience.
Also the tensor structure of $R_{\mu\nu}$ is
left unconstrained by the proof.

\section{Computation of the gluon wave-function renormalization}
\label{app:computations}

The computation of the gluon wave-function renormalization
proceeds by differentiating the reduced flow equation
\Eqref{eq:GonGamma_approx} with respect to $A^{a\mu}(p_1)$
and $A^{b\nu}(-p_2)$, and then setting all fields to 
zero.
This produces a result which is proportional to 
$\delta(p_1-p_2)$. We thus set $p_1=p_2=p$.
The gluon loop, i.e.~the first diagram in
\Figref{fig:nonvanishing}, arises
from the first line in \Eqref{eq:GonGamma_approx}.
This loop comes
in two copies which differ by the propagator
carrying the regulator insertion.
The sum of both reads
\begin{align}
\bs
&\partial_t\Gamma^{(2)}_{A^{a\mu} A^{b\nu}}\Big|_{\text{gluon}}=
\delta^{ab}\delta(p_1-p_2)
\int_q \frac{\partial_t(Z R(q^2))}{P(q^2)^2 P((q+p)^2)}\\
&\bar{g}^2 C_2(G)\delta^{\alpha\beta}\delta^{\lambda\rho}
\hat{\Gamma}^{(3)}_{\lambda\mu\alpha}(-q-p,p,q)\hat{\Gamma}^{(3)}_{\beta\nu\rho}(-q,-p,p+q)\,.
\label{eq:gluonloop}
\es
\end{align}
Here $P$ denotes the regularized inverse gluon propagator
\begin{equation}
P(q^2)= q^2+R(q^2) = q^2 (1+r(q^2))\,,
\end{equation}
and $\hat{\Gamma}^{(3)}$ is the spacetime tensor structure 
of the standard three-gluons vertex
\begin{equation}
\hat{\Gamma}^{(3)}_{\lambda\mu\alpha}(s,p,q)=
\delta_{\lambda\mu}(s-p)_\alpha+\delta_{\mu\alpha}(p-q)_\lambda
+\delta_{\alpha\lambda}(q-s)_\mu\,.
\end{equation}
To extract the correction to the wave-function
renormalization we need to expand \Eqref{eq:gluonloop}
in a Taylor series around $p=0$, and
collect the terms of order $p^2$.
These come into  two forms, proportional to either
$p^2\delta_{\mu\nu}$ or $p_\mu p_\nu$,
and can be organized in
contributions to the transverse or longitudinal
modes.
Furthermore, we focus on the O($\bar{g}^2$)-contribution
and therefore neglect $\partial_t Z$ on the right-hand side.
To evaluate the loop integral, we must choose an explicit  
regulator $R(q^2)$.
Notice however that the final result, being a one-loop
dimensionless counter-term,  is universal, i.e.~regulator-independent.
To analytically perform the integrals we adopt the piecewise linear 
regulator of \Eqref{eq:rlinear}.

The result for the gluon-loop contribution to the gluon anomalous 
dimension is
\begin{align}
\partial_t Z_\T^{\perp} \Big|_{\text{gluon}}&=\partial_t Z_\T^{\parallel} \Big|_{\text{gluon}}=
Z g^2 C_2(G) \frac{19}{96\pi^2} \,,\\
\partial_t Z_\Lo^{\perp} \Big|_{\text{gluon}}&=
\partial_t Z_\Lo^{\parallel} \Big|_{\text{gluon}}=
-Z g^2 C_2(G) \frac{1}{32\pi^2} \,.
\end{align}
This is indeed the universal result for Feynman gauge,
which can be computed also, for instance,
through dimensional regularization~\cite{Peskin:1995ev}.
\begin{figure}[t!]
	\includegraphics[width=\columnwidth]{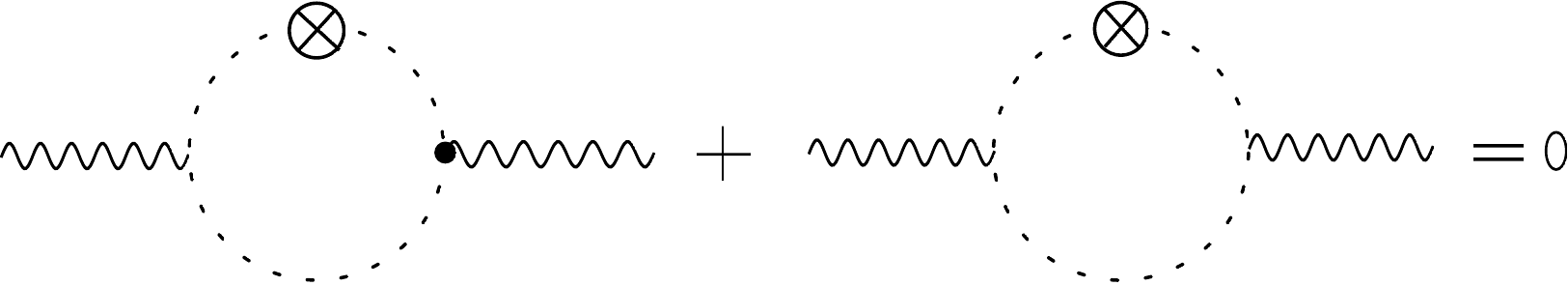}
	%
	\caption{Diagrams which do not contribute to the gluon wave-function renormalization, as they cancel each other. A similar cancellation occurs if one replaces
	the left vertices of both diagrams with $v$-dependent ones (empty circles).}
	\label{fig:vanishing}
\end{figure}
\begin{figure}[t!]
	\includegraphics[width=0.8\columnwidth]{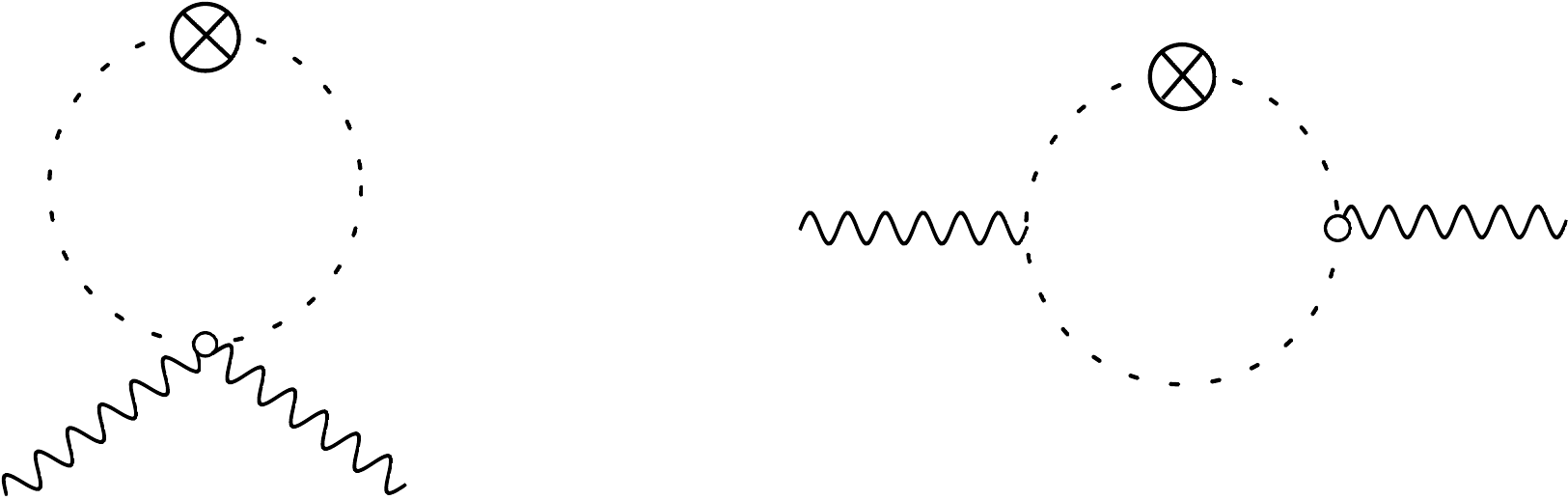}
	%
	\caption{Diagrams which do not contribute to the gluon wave-function renormalization, as they generate a non-quadratic momentum dependence.}
	\label{fig:notcontributing}
\end{figure}
The second line in \Eqref{eq:GonGamma_approx}
cannot contribute, as $\Gamma^{(2)}_{M\Phi}$
contains ghost fields.
Let's then address the ghost contributions to the gluon
wave-function renormalization,
i.e.~the term arising from differentiation of the
third line in \Eqref{eq:GonGamma_approx}.
The second order derivative can be cast in the following
form:
\begin{align}
&\frac{\delta\ }{\delta A^{a\mu}}
\frac{\delta\ }{\delta A^{b\nu}}
\left[
\tGamma^{(2)}_{K^e_\lambda \Phi^j}
\big(\tGamma^{(2)-1}\big)_{\Phi^j(-\bC^d)}
\right]= \nonumber\\
&\bigg\{\Big[\tGamma^{(3)}_{K^c_\lambda  A^{a\mu} \Phi^j}
-\tGamma^{(2)}_{K^e_\lambda \Phi^l}
\big(\tGamma^{(2)-1}\big)_{\Phi^l\Phi^m}
\tGamma^{(3)}_{\Phi^m A^{a\mu}\Phi^j}
\Big]
\label{eq:allghostterms}
\\
&\phantom{XXXXXX}\times
\frac{\delta\ }{\delta A^{b\nu}} \big(\tGamma^{(2)-1}\big)_{\Phi^j(-\bC^d)}
+\left(\begin{pmatrix}
a\\ \mu
\end{pmatrix}\leftrightarrow \begin{pmatrix}
b\\ \nu
\end{pmatrix}\right)\bigg\}
\nonumber\\
&-\tGamma^{(2)}_{K^c_\lambda \Phi^j}
\big(\tGamma^{(2)-1}\big)_{\Phi^j\Phi^i}
\tGamma^{(4)}_{\Phi^i A^a_\mu A^b_\nu \Phi^k}
\big(\tGamma^{(2)-1}\big)_{\Phi^k (-\bC^d)}.
\nonumber
\end{align}
As we must evaluate such derivatives at vanishing fields,
all the $\Phi$'s in this expression have to be either
ghosts or anti-ghosts.
Let us first inspect the $v$-independent
contribution.
In this case the last line of \Eqref{eq:allghostterms}
vanishes, as the usual Lorenz-gauge ghost action
contains no two-ghosts-two-gluons vertex.
The remaining terms correspond to the 
diagrams in \Figref{fig:vanishing}. 
As they carry opposite signs,
they sum up to zero.
In fact,
the $v$-independent terms arising
from the square brackets
in the second line of \Eqref{eq:allghostterms},
once contracted with $\delta^{ed}$,
reduce to $\bar{g} f^{eja} 
(q^2 \delta^\lambda_{\ \, \mu}-q^\lambda q_\mu)/q^2$,
where $q$ is the momentum of $K^a_\lambda(-q)$.
As this expression must be contracted with 
$\partial_t (Z_\gh r_\gh(q^2))\I q_\lambda$,
it drops out of the flow equation.

The $v$-dependent contributions
to the gluon wave-function renormalization
come in four different forms.
Three of them, corresponding to the 
diagrams in \Figref{fig:notcontributing},
are vanishing.
In fact they show no quadratic
term in a Taylor expansion around $p=0$.
The only nonvanishing diagram is
the second one in \Figref{fig:nonvanishing}.
This evaluates to
\begin{align}
&\partial_t\Gamma^{(2)}_{A^{a\mu} A^{b\nu}} \Big|_{\text{ghost}}=\frac{v^a v^b}{2Z^2|v|^4}\int_q 
\frac{{\partial}_t R_{\gh}(q^2) (q+p)^\alpha q^\beta}{P_{\gh}(q^2)^2 P_{\gh}((q+p)^2)}
\nonumber\\
&\times \left(R^{\alpha\mu}(p)R^{\beta\nu}(p) 
+ R^{\alpha\mu}(p)R^{\beta\nu}(q) \right. 
\label{eq:ghostloop}\\
&\phantom{\times 1}\left.+ R^{\alpha\mu}(q+p) R^{\beta\nu}(p) 
+ R^{\alpha\mu}(q+p)R^{\beta\nu}(q)  \right),
\nonumber
\end{align}
where
\begin{align}
R_{\gh}(p^2)&=p^2 r_{\gh}(p^2)\,,\\
P_{\gh}(q^2)&=q^2+R_{\gh}(q^2)\,.
\end{align}
Therefore
\begin{align}
\partial_t Z_\T^{\perp} \Big|_{\text{ghost}}&=0\,,\\
\partial_t Z_\Lo^{\perp} \Big|_{\text{ghost}}&=0\,.
\end{align}
Notice that \Eqref{eq:ghostloop} has already been
multiplied by a factor 2, to account for the
two diagrams with the regulator insertion on different ghost propagators.
To extract the ghost-loop contribution to 
$\partial_t Z_{\T/\Lo}^{\parallel}$
from \Eqref{eq:ghostloop} we first
strip away the factor $\Pi_\parallel^{ab}$.
We next expand \Eqref{eq:ghostloop} in a Taylor series about $p=0$ 
and collect the terms of order $p^2$,
which can be organized in transverse and longitudinal
parts.
Finally we adopt the piecewise linear regulator for both
the gluon and the ghost sector, as in Eqs.~\eqref{eq:equalregulators}
and \eqref{eq:rlinear}.
Setting $p=0$ in the second and third lines of \Eqref{eq:ghostloop}
gives the standard Feynman-gauge result. 
Expanding these very same terms to second order in $p$
provides the corrections due to the regulator-dependent vertices;
see Eqs.~\eqref{eq:finalZTpar} and \eqref{eq:finalZLpar},
and the subsequent discussion
in the main text.

\bibliography{bibliography}

%
%
%
%
%
%
%

\end{document}